\documentclass[aps,pre,twocolumn,showpacs,superscriptaddress]{revtex4-1}
\usepackage{hyperref}
\usepackage[usenames]{color}
\usepackage{color}
\usepackage{colortbl}
\definecolor{linkcolor}{rgb}{0.8,0,0.2}
\definecolor{citecolor}{rgb}{0,0.6,0.2}
\definecolor{urlcolor}{rgb}{0,0,1}
\hypersetup{
    colorlinks, linkcolor={linkcolor},
    citecolor={citecolor}, urlcolor={urlcolor}}
\usepackage{graphicx}
\graphicspath{{Fig/}} % Пути к изображениям
\usepackage{cmap}						% Улучшенный поиск русских слов в полученном pdf-файле
\usepackage[T2A]{fontenc}				% Поддержка русских букв
\usepackage[utf8]{inputenc}				% Кодировка utf8
\usepackage{epstopdf}

\newcommand{\hide}[1]{}

\begin{document}
\title{Chiral Optical Tamm States: Temporal Coupled-Mode Theory}

\author{Ivan V. Timofeev}
\affiliation{Kirensky Institute of Physics, Siberian Branch of the Russian Academy of Sciences, Krasnoyarsk 660036, Russia}
\affiliation{Laboratory for Nonlinear Optics and Spectroscopy, Siberian Federal University, Krasnoyarsk 660041, Russia}
\email{tiv@iph.krasn.ru}
\author{Pavel S. Pankin}
\affiliation{Institute of Engineering Physics and Radio Electronics, Siberian Federal University, Krasnoyarsk 660041, Russia}
\author{Stepan Ya. Vetrov}
\affiliation{Kirensky Institute of Physics, Siberian Branch of the Russian Academy of Sciences, Krasnoyarsk 660036, Russia}
\affiliation{Institute of Engineering Physics and Radio Electronics, Siberian Federal University, Krasnoyarsk 660041, Russia}
\author{Vasily G. Arkhipkin}
\affiliation{Kirensky Institute of Physics, Siberian Branch of the Russian Academy of Sciences, Krasnoyarsk 660036, Russia}
\affiliation{Laboratory for Nonlinear Optics and Spectroscopy, Siberian Federal University, Krasnoyarsk 660041, Russia}
\author{Wei Lee}
\affiliation{Institute of Imaging and Biomedical Photonics, College of Photonics, National Chiao Tung University, Guiren Dist., Tainan 71150, Taiwan}
\author{Victor Ya. Zyryanov}
\affiliation{Kirensky Institute of Physics, Siberian Branch of the Russian Academy of Sciences, Krasnoyarsk 660036, Russia}

\date{\today}

\begin{abstract}
The chiral optical Tamm state (COTS) is a special localized state at the interface of a handedness-preserving mirror and a structurally chiral medium such as a cholesteric liquid crystal or a chiral sculptured thin film. 
The spectral behavior of COTS, observed as reflection resonances, is described by the temporal coupled-mode theory. 
Mode coupling is different for two circular light polarizations because COTS has a helix structure replicating that of the cholesteric. 
The mode coupling for co-handed circularly polarized light exponentially attenuates with the cholesteric layer thickness since the COTS frequency falls into the stop band. 
Cross-handed circularly polarized light freely goes through the cholesteric layer and can excite COTS when reflected from the handedness-preserving mirror. 
The coupling in this case is proportional to anisotropy of the cholesteric and theoretically it is only anisotropy of magnetic permittivity that can ultimately cancel this coupling. 
These two couplings being equal results in a polarization crossover (the Kopp--Genack effect) for which a linear polarization is optimal to excite COTS. 
The corresponding cholesteric thickness and scattering matrix for COTS are generally described by simple expressions.

\end{abstract}

\maketitle

\section*{Introduction}
\begin{figure}[b]%[htbp]
\center{\includegraphics[scale=.75]{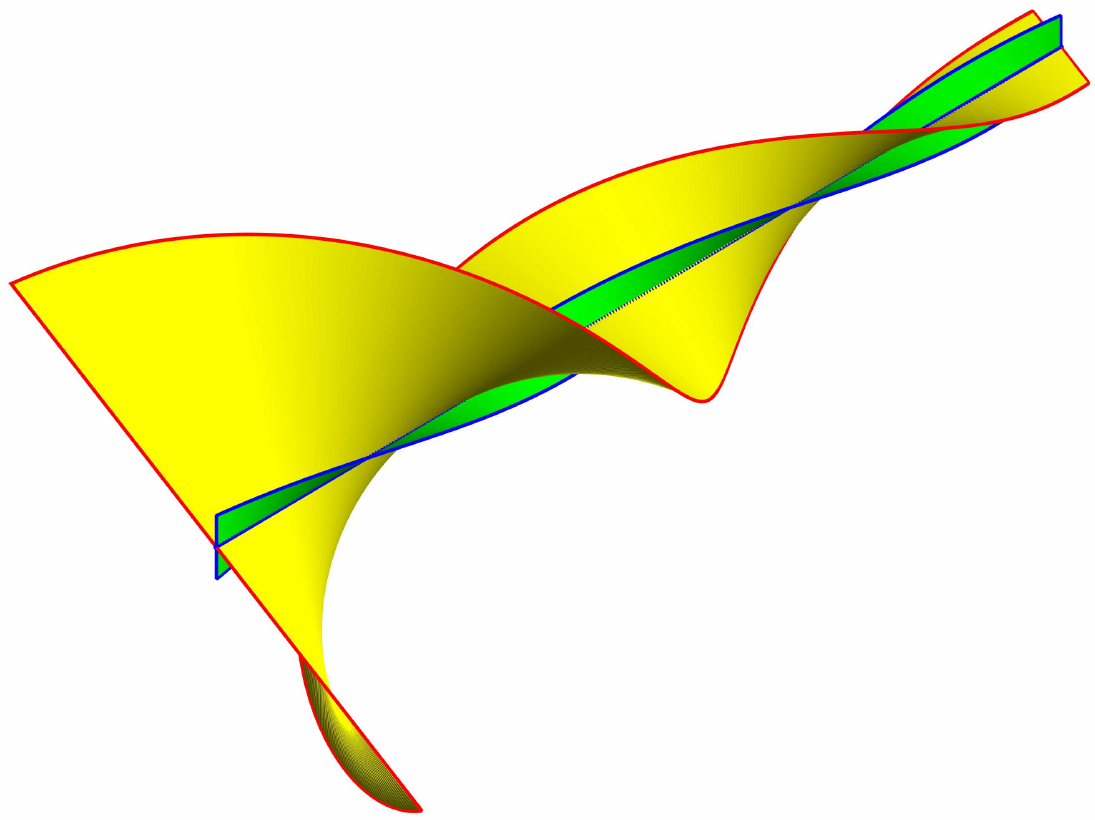}}
\caption{%
Circular Bragg diffraction forming COTS. 
The cholesteric director is shown in blue and green, the electric field is in red and yellow. 
The angle between them does not change with depth}
\label{Fig:OTS}
\end{figure}

Matter tends to order thus forming crystals. 
Orientation alignment is the preferred order in liquid crystals. 
Due to its cyclic nature, it can generate an echo of a translational order in chiral superlattices of cholesteric liquid crystal. 
Under diffraction, in cholesteric crystals the helical structure of the matter is rendered to the field because the co-handed circularly polarized wave diffracts (Fig.~\ref{Fig:OTS}) while the cross-handed circularly polarized wave travels virtually unaffected. 
This phenomenon is referred to as selective reflection \cite{Belyakov1992b} or, alternatively, Bragg circular diffraction for electromagnetic and acoustic waves \cite{Faryad_Lakhtakia2014rv}. 

The results obtained are readily generalized for any material with a helix-like response, including widely tunable heliconical structures \cite{Xiang_Lavrentovich2015}. 
The selective reflection obstructs observation of localized states when the order at the interface or at the structural defect is disturbed. 
This is paid off by a simpler description of the states, which is an advantage offered by smooth helix symmetry \cite{Belyakov1992b,Avendano_Oldano2005} as opposed to discrete translational symmetry of crystals. 
The structural defect is conventionally represented as a cavity confined by mirrors, where the role of mirrors is played by the Bragg grating. 
This counterpart of a Fabry--Perot resonator possesses a number of defect modes. 
The latters are localized optical states normally corresponding to the whole number of halfwaves accommodated in the cavity. 
There are a number of distinctive polarization features \cite{Haus1976,Lakhtakia1999_Defect,Yang1999,Shabanov2004,Gevorgyan2006,Belyakov2011,Kiselev2014} associated with the chiral defect applications \cite{Faryad_Lakhtakia2014rv,Hsiao2014,Rodarte2015,LinJiaDe2015,Kesaev2017}, among which the twist defect, lacking an intermediate layer and having zero thickness, is the most prominent \cite{Hodgkinson2000_Twist,Kopp2002}. 
Strict theoretical \cite{Oldano2003,Becchi_Oldano2004,Avendano_Oldano2005,Schmidtke2003_EPJ} and experimental \cite{Hodgkinson2000_Twist,Schmidtke2003_PRL,Ozaki2003_CLC_laser,Shibaev2002_CLC_laser} studies of the phenomenon gave rise to a discussion on polarization and relaxation time of the localized state \cite{Oldano2003_comment,Kopp2003_comment}. 
Theoretically, infinite relaxation time is only possible if there is anisotropy of magnetic permittivity \cite{Avendano_Oldano2005,Gevorgyan_Rafayelyan2013}. 
Otherwise it appears impossible to simultaneously match electric and magnetic field strengths at the interface. 
Note that infinitely increasing the cholesteric thickness does not provide infinite improvement of the quality factor. 
The quality factor saturates with increasing cholesteric thickness and circular polarization of the transmitted light changes from co-handed to cross-handed. 
This polarization crossover \cite{Kopp2002} has been called the Kopp--Genack effect \cite{McCall_Hodgkinson2015bk}. 
Unlike the Fabry--Perot resonator, the twist defect generates a single localized state. 
The spatial field distribution curve in this case has no flat top and consists of two waves exponentially descending in opposite directions. 
This resembles a surface wave with the only difference that for the twist defect there is no limitation on the angle of incidence of the excitation wave, and surface waves at the cholesteric-isotropic dielectric interface are only observed at the angles ensuring complete internal reflection \cite{Belyakov1992_Shilina}. 

There exists a surface wave beyond the restriction of complete internal reflection. 
It is known as an optical Tamm state (OTS) \cite{Kavokin2005,Vinogradov2010rv,Iorsh2012,Afinogenov2013,Auguie_Fainstein2015,ChenKuoping2015,ChenKuoping2016}, which is similar to the Tamm state of electrons at the superlattice interface. 
The dispersion of OTS lies outside the light cone given by $k=\omega/c$ \cite{Kaliteevski2007a}. 
Such a state can be excited even perpendicular to the surface without energy transfer along the surface, which is an advantage for various applications. 
A question naturally arises if there is an OTS at the cholesteric-metal interface when the light is incident perpendicularly. 
When dealing with this problem, one should bear in mind two things: first, semitransparence of the cholesteric due to the circular Bragg diffraction and, second, polarization change caused by alternating circular polarizations, namely: reflection from cholesteric does not change the handedness of circular polarization whereas reflection from metal does \cite{Abdulhalim2006}. 
This alternation acts like traffic lights: a co-handed circular polarized wave is not allowed to go through the cholesteric until after it has been twice reflected from the metal \cite{Timofeev2013}. 
The polarization match at the interface between chiral and nonchiral mirrors can be achieved by adding an extra anisotropic layer. 
This layer produces a set of localized states with nontrivial polarization properties \cite{Pyatnov2014m_OL,Pyatnov2016m_JOpt,Pyatnov2017m_LC}. 
Various combinations involving two mirrors have been extensively studied and offered for practical applications 
\cite{Zhuang_Patel1999,Abdulhalim2006},
including less-than-one-pitch chiral layer 
\cite{Isaacs_Abdulhalim2014,Hsiao2011,Timofeev2012,Timofeev2016t,Gunyakov2016_JQSRT,HuangKuanChung2016m}.
The closer to the mirror, the higher the energy density of the states may become, but still the states are not localized at surface but within the bulk of the extra layer. 
It is possible to do without an extra layer provided a handedness-preserving mirror (HPM) is used \cite{Plum_Zheludev2015,Fedotov_Zheludev2005,Kopp2003_comment,Rudakova2017tm_BRAS}. 
HPM maintains not only the handedness but also the ellipticity magnitude upon reflection, therefore, such a mirror is also referred to as a polarization-preserving anisotropic mirror \cite{Rudakova2017tm_BRAS}. 
Also HPM can be defined as a reflector with the effect of half-wave phase plate \cite{Kildishev2015}.
A localized state at the HPM/choleric interface was described in the low anisotropy approximation of cholesteric crystal and was called a chiral optical Tamm state (COTS) \cite{Timofeev2016t}. 

In this paper we seek to answer the question whether this state is possible in principle with an ideal HPM and a semi-infinite non-absorbing cholesteric layer having finite anisotropy. 
A detailed description is given of the simplest case when the electric and magnetic anisotropies are identical. 
Two types of deviation are considered: the lack of magnetic anisotropy and the finite thickness of the cholesteric layer. 
The ideal state here becomes a resonance with a finite quality factor and relaxation time. 

\subsection*{A method to describe spectral peaks}
The spectra of interest and the field distribution are conveniently described by the Berreman formalism \cite{Berreman1972,Palto2001}. 
For a normal incidence on cholesteric there is an uncomplicated exact solution \cite{Oseen1933,DeVries1951,Kats1971,Nityananda1973}. 
By matching the tangential strengths at the cholesteric interface we can write down general closed-form equations \cite{Oldano2003,Becchi_Oldano2004,Avendano_Oldano2005,Schmidtke2003_EPJ}. 
For the sake of simplicity and clarity, we additionally use an approximate analytical method -- the temporal coupled-mode theory (TCMT), or the theory of coupled modes in the time domain \cite{Haus1984bk,Manolatou2002bk,Joannopoulos2008bk}. 
TCMT provides an instrument to describe the field in coupled resonators where there is coupling between the resonator and the waveguide. 
The spatial structure of the localized mode here is not involved. 
It is the complex amplitude of this mode and its time derivative that matter for this theory. 
TCMT is a popular approach for dealing with stationary processes where the time derivative is zero. 
The first word, `temporal',  in the term \cite{FanShanhui2003} can be treated as historically coined. 
Essentially the same method is employed to describe open resonators \cite{Xu_Yariv2000,Bliokh2008rv} and it goes back to the Lippmann and Schwinger's solution of the scattering problem in quantum mechanics \cite{Lippmann1950}. 
This method is not to be confused with the theory of coupled waves \cite{Kogelnik1969,Yariv1973_CMT}, or the spatial theory of coupled modes widely used in the optics of cholesterics \cite{Belyakov1974,Belyakov1979,Belyakov1992_Shilina,Belyakov1992t_ShilinaScalar,McCall2000,Wang2005}. 
While both approaches rely on the concept of coupled modes \cite{Pierce1954}, the latter theory deals with coupled amplitudes of propagating waves, leaving the amplitude of the resonator mode outside the scope of consideration.

\section*{Model}
A sketch of the cholesteric interface is shown in Fig.~\ref{Fig:OTS}. 
The cholesteric helical axis is perpendicular to the mirror surface. 
In other words, the cholesteric director, i.e. 
the unit vector of the preferred orientation of molecules, is constant in the interface-parallel cross-sections and it uniformly rotates with increasing distance from the interface. 
Near the interface, a chiral optical Tamm state (COTS) is possible, described in \cite{Timofeev2016t} in the limit of low cholesteric anisotropy. 
This state can be represented as a superposition of two circularly polarized counter-propagating co-handed waves with their strengths rotating in time in opposite directions. 
The resultant polarization is linear at each point of space, and the plane of polarization uniformly rotates together with the cholesteric director as the distance from the interface increases. 
The amplitude exponentially drops without standing-wave nodes and antinodes. 
The field structure resembles a twisted onion dome of Saint Basil's Cathedral in Moscow or Dutch Renaissance style in Copenhagen Stock Exchange dome. 

\begin{figure}[b]%[htbp]
\center{\includegraphics[scale=1]{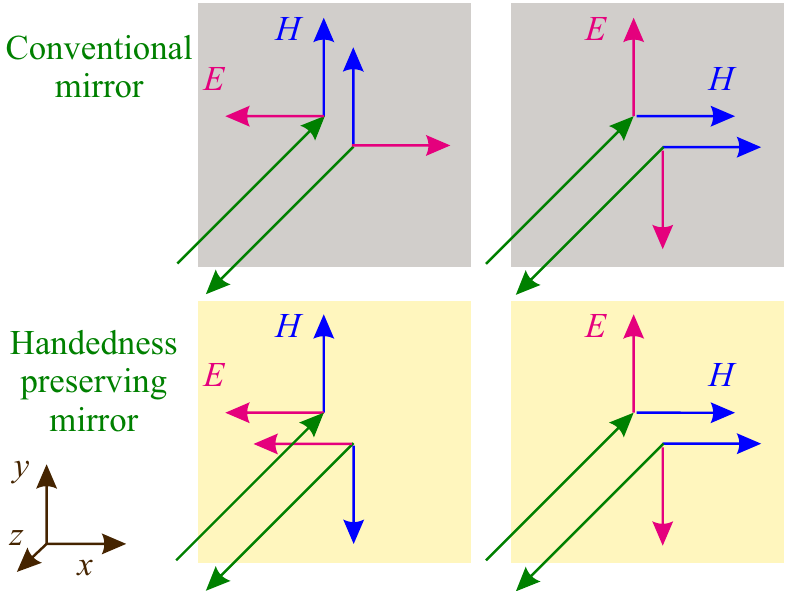}}
\caption{%
HPM and a conventional mirror. 
The $x$-polarized light reflection is different. For HPM the reflected electric strength $E_x$ preserves its phase and the phase of the magnetic strength $H_y$ alters instead.}
\label{Fig:HPM}
\end{figure}
Fig.~\ref{Fig:HPM} compares HPM to a conventional mirror. 
In a conventional metallic mirror, the electric field, when reflected, reverses its phase whereas the magnetic field does not. 
A reverse situation is possible, in which case the mirror is called a magnetic mirror. 
HPM combines the electric and magnetic types of reflection \cite{Fedotov_Zheludev2005}. 
Let $y$ be such an axis in the HPM plane that an electric field linearly polarized along that axis changes its phase after reflection. 
The orthogonal axis $x$ corresponds to magnetic reflection. 
It is important that the magnetic field is orthogonal to the electric one, and hence to the $x$ axis, and is directed along $y$. 
In other words, the electric and magnetic field components directed along $y$ are subject to a phase jump and yield a node, i.e. 
zero intensity. 
It is only $x$-components of the field that remain non-zero.

\subsection*{Maxwell equations in the basis associated with the cholesteric director}
Here we limit ourselves to the case of normal light incidence. 
For $\mu=1$=1, the Maxwell equation for a wave propagating along the helix axis $z$ can be written as
	\begin{equation}
	\frac{\partial^2 \vec{E}}{\partial z^2} = 
	\frac{\hat{\epsilon}}{c^2}
	\frac{\partial^2 \vec{E}}{\partial t^2}. 
	\end{equation}
The wave is described by a vector $\vec{E}$ of complex amplitudes for the electric field components in the orthogonal directions $x$ and $y$. 
Projection of the dielectric permittivity tensor $\hat{\epsilon}$ on the plane $x,y$ at the depth of cholesteric $z$ is given by
	\begin{equation}
	\hat{\epsilon} = \left[\begin{array} {cc}
	\epsilon_e \cos^2 \tilde{\phi} + \epsilon_o \sin^2 \tilde{\phi} & \sin 2\tilde{\phi} \;  (\epsilon_e-\epsilon_o)/2 \\ 
	\sin 2\tilde{\phi} \; (\epsilon_e-\epsilon_o)/2 & \epsilon_e \sin^2 \tilde{\phi} + \epsilon_o \cos^2 \tilde{\phi} 
	% \epsilon_o \cos^2 \tilde{\phi} + \epsilon_e \sin^2 \tilde{\phi} & \sin 2\tilde{\phi} \;  (\epsilon_e-\epsilon_o)/2 \\ 
	% \sin 2\tilde{\phi} \; (\epsilon_e-\epsilon_o)/2 & \epsilon_o \sin^2 \tilde{\phi} + \epsilon_e \cos^2 \tilde{\phi} 
	\end{array} \right]. 
	\end{equation}
Here the optical axis coinciding with the cholesteric director is given by the twist angle $\tilde{\phi}(z) = 2\pi z /p +\phi$, which is measured from the $x$ axis toward the $y$ axis; $p$ is the helix pitch. 
The positive pitch refers to a right-handed helix, the negative pitch refers to a left-handed one. 
Taking into account magnetic permittivity $\hat{\mu}_{xy}$ makes an explicit form of the magnetic strength $\vec{H}$ expression more preferred and increases the field vector dimensionality from 2 to 4. 

\[
	\vec{J} = [E_x,H_y,E_y,-H_x]^T. 
	\]
Consequently, the order of the differential equation goes down from second to first. 
Let us assume the principal axes of the magnetic and electric permittivity tensors to coincide. 
On this assumption, we can use the orthonormal basis $u,v,z$ uniformly rotating together with the cholesteric director so that the $u$ axis always goes along the director:
\[
	\vec{J}_R = [E_u,H_v,E_v,-H_u]^T. 
	\]
By the Berreman method \cite{Berreman1972}, Maxwell equations in a stationary case have the form:
	\begin{equation}
	\frac{\partial \vec{J}_R}{\partial \tilde{z}} = 
	i \hat{T}\vec{J}_R. 
	\label{eq:Berreman}
	\end{equation}
A differential transfer matrix for rotating basis is formulated in \cite{Oldano2003,Avendano_Oldano2005} and it can be reduced as follows: 
	\begin{equation}
	\hat{T} = \left[\begin{array} {cccc}
	0 & \mu_o & \tilde{\lambda} & 0 \\ 
	\epsilon_e & 0 & 0 & \tilde{\lambda} \\ 
	\tilde{\lambda} & 0 & 0 & \mu_e \\ 
	0 & \tilde{\lambda} & \epsilon_o & 0 \\ 
	\end{array} \right],
	\label{eq:TM}
	\end{equation}
where $\tilde{\lambda} = \lambda_0/p =  2 \pi c / \omega p$ is the nondimensional wavelength, $\tilde{z} = z \omega/c=2\pi z/ \lambda_0$ is the nondimensional coordinate. 
There are different units for electric and magnetic strengths in the SI system, therefore they have to be normalized via the vacuum impedance $Z_0=E/H=\sqrt{\mu_0/\epsilon_0}$. 

\subsection*{Solution without low anisotropy approximation}  
Four normal waves correspond to Eq.(\ref{eq:Berreman}). 
They are determined by the eigenvalues of the $\hat{T}$ matrix These eigenvalues have the sense of refractive indices $n$ and the respective eigenvectors of the $\hat{T}$ matrix have the sense of polarizations $\vec{J_0}$. 
Based on the $z$-axis reversal symmetry $z \to -z$, these four normal waves can be classified as two pairs of counter-directed waves. 
In each pair, the wave with a larger refractive index has a lower phase velocity. 
We refer to this wave as a slow wave. 
The other wave is called a fast wave:
	\begin{equation}
	\vec{J}_{R0} = \vec{J}^\pm_{s,f}\exp(\pm i n_{s,f}\tilde{z}). 
	\label{eq:J_eigen}
	\end{equation}
Substituting the solution for $\vec{J}^+_{s,f}$ into Eq. 
(\ref{eq:Berreman}) yields:
	\begin{equation}
	i [\hat{T} - n_{s,f} \hat{I}] \vec{J}_{s,f} = 0,
	\label{eq:eigen}
	\end{equation}
where $\hat{I}$ is the unit matrix, the index $+$ in $\vec{J}^+_{s,f}$ has been omitted. 
The refractive indices are as follows
	\begin{equation}
	n_{s,f}^2 = \tilde{\lambda}^2 + \overline{\epsilon\mu}
	\pm \sqrt{4\bar{\epsilon}\bar{\mu}\tilde{\lambda}^2+d_{\epsilon\mu}^2},
	\label{eq:eigenvalue}
	\end{equation}
where $d_{\epsilon\mu}=(\epsilon_e\mu_o-\epsilon_o\mu_e)/2$ is the antisymmetry coefficient of permittivities and the overbar means the arithmetic mean over ordinary and extraordinary permittivities:
	\begin{equation}
	\bar{\epsilon}=\frac{\epsilon_e+\epsilon_o}{2},
	\bar{\mu} = \frac{\mu_o+\mu_e}{2},
	\overline{\epsilon\mu}=\overline{\epsilon_{eo}\mu_{oe}}=\frac{\epsilon_e\mu_o+\epsilon_o\mu_e}{2}. 
	% \label{eq:eigenvalue}
	\end{equation}
Scale invariance of Maxwell equations (\ref{eq:Berreman}) and normalization of material parameters (see Supplement in \cite{Avendano_Oldano2005}) reduces the structure to two crucial parameters: electric and magnetic anisotropies
	\begin{equation}
	\delta_\epsilon = \frac{\epsilon_e-\epsilon_o}{\epsilon_e+\epsilon_o},
	\delta_\mu = \frac{\mu_e-\mu_o}{\mu_e+\mu_o}. 
	\end{equation}
\subsection*{The case of equal anisotropies, $\delta_\epsilon = \delta_\mu$}
Consider the case of equal anisotropies $\delta = \delta_\epsilon=\delta_\mu$. 
The clarity and fineness of this single-parametric set of structures should compensate the difficulty of their physical realization in the optical range for the reader \cite{Gevorgyan_Rafayelyan2013}. 
The wavelength parameterizes the set of differential transfer matrixes (\ref{eq:TM}) the way eccentricity does to a set of conical cross-sections. 
The dispersion equation simplifies due to the symmetry of permittivities $d_{\epsilon\mu}=0$
	\begin{equation}
	n_{s,f}^2 = (\tilde{\lambda} \pm \sqrt{\bar{\epsilon}\bar{\mu}})^2-(\bar{\epsilon}\bar{\mu}-\overline{\epsilon\mu}),
	% \label{eq:eigenvalue_symmetric}
	\end{equation}

\begin{figure}[htbp]
\includegraphics[scale=1]{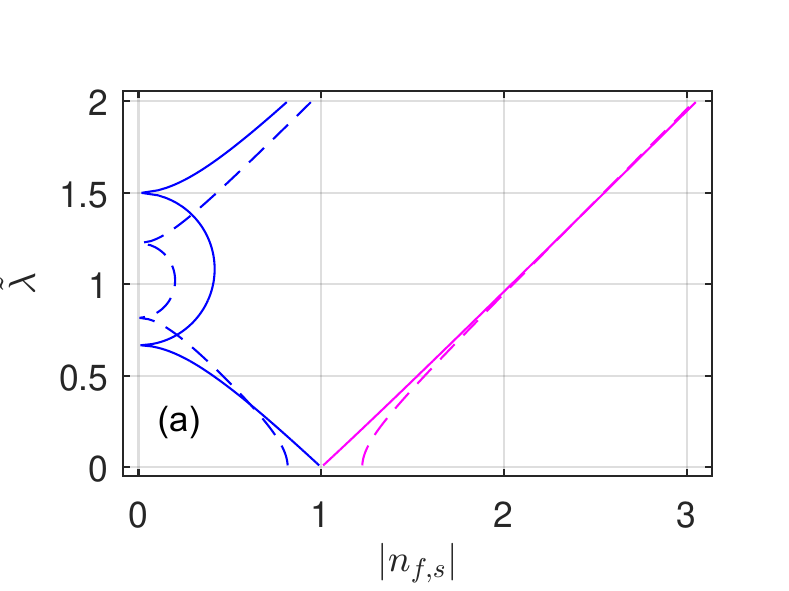}
\includegraphics[scale=1]{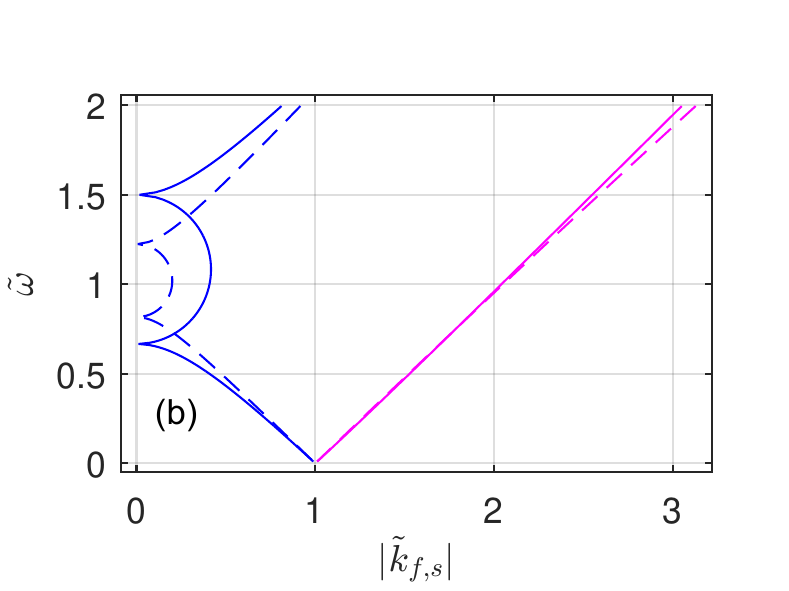}
\caption{%
Symmetry of dispersion curves. 
(\textbf{a}) Wavelength $\tilde{\lambda}$ as a function of the refractive index $|n_{f,s}|$. 
(\textbf{b}) Frequency $\tilde{\omega} = 1 / \tilde{\lambda}$ as a function of the wave number $|\tilde{k}_{f,s}|$. 
The blue curve is the fast wave, the purple curve -- slow wave. 
The solid curve is drawn for $\epsilon_e=\mu_e=3/2, \epsilon_o=\mu_o=2/3$, $\delta_\epsilon=\delta_\mu$. 
The semicircle refers to a diffracting wave for which the refractive index acquires purely imaginary values $|n_{f}|=Im(n_{f})$. 
The Mauguin regime, $\tilde{\lambda} \ll 1$, is equivalent to homogenization $\omega \ll 2\pi$ according to dispersion equations (\ref{eq:TM}). 
The dashed line is  $\epsilon_o=2/3, \epsilon_e=3/2, \mu_e=\mu_o=1$. 
The symmetry is violated
}
\label{Fig:BandStructure}
\end{figure}

Without further prejudice to the generality, we assume the normalization $\bar{\mu} = \bar{\epsilon}$, $\overline{\epsilon\mu} = 1$. 
Then $\sqrt{\bar{\epsilon}\bar{\mu}} = \bar{\epsilon} \ge 1$. 
In other words, the permittivity is normalized to the geometric mean of $\epsilon_o$ and $\epsilon_e$: $\bar{\epsilon}_g = \sqrt{\epsilon_o \epsilon_e}=1$, the arithmetic mean being not less than unity $\bar{\epsilon} \ge 1$. 
The second term in the right-hand part of the dispersion equation becomes squared anisotropy $\bar{\epsilon}\bar{\mu}-\overline{\epsilon\mu} = \delta^2$. 
Anisotropy here acquires the meaning of the standard deviation of permittivities:
	\begin{equation}
	n_{s,f}^2 = (\tilde{\lambda} \pm \bar{\epsilon})^2-\delta^2. 
	\label{eq:eigenvalue_symmetric}
	\end{equation}
Unlike the parabolic approximation typical of periodic media, the dispersion curves have a hyperbola shape, except for the stop band where they have the shape of a circle. 
Dispersion equation (\ref{eq:eigenvalue_symmetric}) can be written for the refractive index as well as for the wave vector
	\begin{eqnarray}
	n_{s,f}^2 = \tilde{\lambda}^2 \pm 2\bar{\epsilon}\tilde{\lambda} + 1,
	\nonumber \\
	\tilde{k}_{s,f}^2 = n_{s,f}^2/\tilde{\lambda}^2 = \tilde{\omega}^2 \pm 2\bar{\epsilon}\tilde{\omega} + 1. 
	\label{eq:eigenvalue_symmetric_wk}
	\end{eqnarray}
This symmetry of $\tilde{\lambda}(n)$ and $\tilde{\omega}(\tilde{k}) = 1 / \tilde{\lambda}$ dispersions indicates the symmetry of the longwave and shortwave limits. 
In the longwave limit, the medium is homogenized and becomes isotropic. 
The negative optical activity ceases as the situation comes close to the static field case. 
In the shortwave limit, positive optical activity is supported by the Mauguin waveguide regime (Fig.~\ref{Fig:BandStructure}). 
This symmetry is destroyed when the anisotropies become unequal $\delta_\epsilon \neq \delta_\mu$. 
The ordinary and extraordinary waves can then be distinguished in the high-frequency limit equivalent to helix untwisting. 
Next we focus on the circular Bragg diffraction, when $(\tilde{\lambda} - \sqrt{\bar{\epsilon}\bar{\mu}})^2 \le \delta^2$ and the refractive index for the fast wave $n_f$ acquires purely imaginary values. 
This is the case when the phase velocity becomes infinite and the group velocity -- meaningless. 
It would be reasonable here to write down dispersion equation (\ref{eq:eigenvalue_symmetric}) in the form of a trigonometric identity where some angle $\chi \in [0,\pi/2]$ acts instead of the wavelength, on the physical meaning of which we will dwell later. 
	\begin{equation}
	- \delta^2 \sin^2 2\chi = \delta^2 \cos^2 2\chi - \delta^2. 
	\end{equation}
This means that at wavelength
	\begin{equation}
	\tilde{\lambda}_0 = \bar{\epsilon}+ \delta \cos 2\chi,
	\label{eq:angle_dispersion_symmetric}
	\end{equation}
the refractive index of the fast wave $n_f = i \delta \sin 2\chi$ describes full reflection in the cholesteric bulk. 
This wave is conventionally called a diffracting wave \cite{Belyakov1992b} and its polarization is derived from nontrivial solvability of Eq. 
(\ref{eq:eigenvalue_symmetric}). 
Subtracting the refractive index from transfer matrix (\ref{eq:eigen}) yields the following identity:
\begin{widetext}
	\begin{eqnarray*}
	[\hat{T}-n_f \hat{I}] \vec{J}_f =
	\delta \left[\begin{array} {cccc}
	-i \sin 2\chi 		& \Delta-\bar{\epsilon} 		& - i \Delta - i \cos 2\chi 		& 0 \\ 
	\Delta+\bar{\epsilon} 		& -i \sin 2\chi  		& 0 		& - i \Delta - i \cos 2\chi \\ 
	i \Delta + i \cos 2\chi 		& 0 		& -i \sin 2\chi  		& \Delta+\bar{\epsilon} \\ 
	0 		& i \Delta + i \cos 2\chi 		& \Delta-\bar{\epsilon} 		& -i \sin 2\chi  \\ 
	\end{array} \right] 
	\left[\begin{array} {r}
	\cos \chi \\ 
	- i \sin \chi \\ 
	- \sin \chi \\ 
	- i \cos \chi \\ 
	\end{array} \right] \equiv 0,
	% \label{eq:TM}
	\end{eqnarray*}
\end{widetext}
where $\Delta = \bar{\epsilon}/\delta$. 
We deal here with a manifest symmetry with respect to exchange between electric and magnetic fields and permittivities, $E_u/E_v=(-H_v/H_u)^*$. 
The electric and magnetic permittivities being equal is equivalent to the Umov-Pointing vector going to zero the same way as for conventional standing-wave nodes and antinodes. 
The electric and magnetic fields are linearly polarized at the same angle $-\chi$ to the cholesteric director. 
This direction coincides with the $x$ axis for $\chi = \phi = \tilde{\phi}(z=0)$. 
Surface conditions for an ideal HPM are exactly the same, i.e. 
both the electric field and the magnetic field are linearly polarized along the magnetic axis of the mirror (Fig.~\ref{Fig:HPM}). 
Hence the derived expression provides an exact description of COTS for a finite anisotropy. 

\subsection*{The case of unequal anisotropies, $\delta_\epsilon \neq \delta_\mu$}
In a general case, we denote directions of electric and magnetic polarizations by the angles $\chi_E$ and $\chi_H$. 
Generalized nontrivial solution (\ref{eq:eigenvalue_symmetric_wk}) provides eigenvectors of the transfer matrix (\ref{eq:eigen}) and exact solutions for the angles \cite{Becchi_Oldano2004}:
	\begin{eqnarray}
	&&\tan{\chi_E} =
	\frac{E_u}{E_v} =
	- \tilde{\lambda}_0 n_f \frac{\mu_e+\mu_o}{\epsilon_e \mu_e \mu_o - n_f^2 \mu_o - \tilde{\lambda}_0^2 \mu_e},
		\nonumber  \\
	&&-\cot{\chi_H} =
	\frac{H_v}{-H_u} =
	- \tilde{\lambda}_0 n_f \frac{\epsilon_e+\epsilon_o}{\mu_e \epsilon_e \epsilon_o - n_f^2 \epsilon_o - \tilde{\lambda}_0^2 \epsilon_e},
		\nonumber \\
	&&\psi = \chi_E - \chi_H \neq 0,
		\nonumber \\
	&&\bar{\chi} = (\chi_E + \chi_H)/2,
	\end{eqnarray}
where the refractive index $n_f$ should be borrowed from dispersion equation (\ref{eq:TM}). 
It is evident that the polarizations are linear for the purely imaginary refractive index $n_f$ whereas for the real refractive index $n_s$ they are almost circular in the stop band. 
Dispersion equation (\ref{eq:eigenvalue}) can be generalized as:
	\begin{equation}
	\tilde{\lambda}_0 \approx \sqrt{\bar{\epsilon}\bar{\mu}}+\sqrt{\bar{\epsilon}\bar{\mu}-\overline{\epsilon\mu}} \cos 2\bar{\chi}. 
	\label{eq:angle_dispersion}
	\end{equation}
For a cholesteric without magnetic anisotropy, $\mu=1$, we have:
	\begin{equation}
	\tilde{\lambda}_0 \approx {\sqrt{\bar{\epsilon}}+\sqrt{\bar{\epsilon}-1} \cos 2\bar{\chi}}. 
	\label{eq:angle_dispersion_mu1}
	\end{equation}
The eigenfrequency is:
	\begin{equation}
	\tilde{\omega}_0 \approx \frac{1}{\sqrt{\bar{\epsilon}}+\sqrt{\bar{\epsilon}-1} \cos 2\bar{\chi}}. 
	\label{eq:angle_dispersion_mu1_omega}
	\end{equation}
This expression is more exact compared to the earlier obtained approximation (Eq.(\ref{eq:angle_dispersion_symmetric}) in \cite{Timofeev2016t}) and still there remains approximation. 
We have to use a unified angle $\bar{\chi}$ between the cholesteric optical axes and HPM hence the electric and magnetic polarizations are directed differently in a general case. 
This does not meet the condition on the HPM surface. 
The non-zero angle of polarization mismatch $\psi$ introduces a new non-local COTS component \cite{Becchi_Oldano2004}. 
This results in the state becoming a leaking mode (or resonance) and acquiring finite relaxation time $\tau_\psi$. 

\subsection*{COTS relaxation time}

By definition, the time of relaxation of a vibrational state is the ratio between the stored energy, $\mathcal{E}$, and the lost power, $P$, taken with a positive sign:
	\begin{equation}
	\frac{\tau}{2} 
	= \frac{\mathcal{E}}{P}
	= \frac{\mbox{Energy stored}}{\mbox{Power of leakage}}.
	\label{eq:tau_def}
	\end{equation}
Here $\tau$ is the amplitude relaxation time, which is twice the energy relaxation time.
Since field does not penetrate into depth of an ideal HPM, the whole stored energy is the energy of diffracting waves inside the cholesteric layer.
Let us find this energy by integrating its density over space $W = [\vec{E}\vec{D}+\vec{H}\vec{B}]/8\pi$, in Gaussian units. 
The period-averaged densities of electric and magnetic energy components are equal because 
$ \overline{\vec{E}\vec{D}} = \overline{\vec{H}\vec{B}}$. 
Moreover $\overline{|E^2|} = |E_0^2|/2$, where $E_{0}$ is the wave amplitude. 
Therefore, $\bar{W} = \bar{\epsilon} |E_{0}^2|/8 \pi$. A standing wave in the layer is formed by two waves travelling in opposite directions. 
Their constructive interference is compensated by a destructive one, and their energy densities add up. 
In the defect layer, $\bar{W}_0 = \bar{\epsilon} |E_0^2|/4 \pi$.
Integration yelds
	\begin{equation}
	\mathcal{E} = \int_{0}^{\infty} \bar{W}_0 \exp(2 i n_f \cdot 2\pi z/\lambda_0) dz = \frac{\bar{W}_0 \lambda_0}{4\pi |n_f|}.
	\end{equation}
With account of Eq.~(\ref{eq:eigenvalue},\ref{eq:angle_dispersion_mu1}) for $\delta_\mu=0$
and $\bar{\epsilon} \approx 1$,
	\begin{equation}
	|n_f| = (\delta_\epsilon/2) \sin {2\bar{\chi}}, \,\;
	\lambda_f = {\lambda_0}/{|n_f|}
	\end{equation}
Here $\lambda_0 = \tilde{\lambda}_0 p$ is given by (\ref{eq:angle_dispersion}). 
The power of leakage is proportional to the flow velocity $c / \sqrt{\bar{\epsilon}}$ and the energy density $\bar{W}_\psi = \bar{\epsilon} |E_\psi^2|/8 \pi$ carried away by the travelling wave. 
The strength is governed by the boundary conditions:
	\begin{equation}
	E_\psi = 2 E_0 \sin (\psi/2). 
	\label{eq:J_psi}
	\end{equation}
We finally have \cite{Becchi_Oldano2004}
	\begin{equation}
	\tau_\psi = 
	\frac{\lambda_f}{4\pi c}
	\frac{1}{\sin^2 (\psi/2)}. 
	\label{eq:t_psi}
	\end{equation}
Consider a cholesteric layer of finite thickness $L$ embedded in a medium with permittivity $\bar{\epsilon}_g = \sqrt{\epsilon_o \epsilon_e}$. 
Instead of (\ref{eq:J_psi}), the power of leakage at the edge of the cholesteric is given by:
	\begin{equation}
	E_L = 2 E_0 \exp \left( - 
	\frac{2\pi L}{\lambda_f}
	% \frac{\pi L \delta_\epsilon \sin 2\bar{\chi}}{\lambda_0}
	 \right). 
	\end{equation}
Note that because of cholesteric boundary condition $E_L$ is twice higher than just exponentially decreasing by Eq.~(\ref{eq:J_eigen}). Then the corresponding relaxation time is
	\begin{equation}
	\tau_L = 
	\frac{\lambda_f}{4\pi c}
	% \frac{\lambda_0}{2\pi c \delta_\epsilon \sin 2\bar{\chi}}
	\exp \left(
	\frac{4\pi L}{\lambda_f}
	% \frac{2\pi L \delta_\epsilon \sin 2\bar{\chi}}{\lambda_0}
	\right),
	\label{eq:t_L}
	\end{equation}
which agrees with the expressions obtained in \cite{Becchi_Oldano2004,Belyakov2011}. The last formula is also directly applicable when $\bar{\epsilon}_g \neq 1$, assuming $L$ being optical density $L = L_0 \sqrt{\bar{\epsilon}_g}$

\subsection*{Temporal coupled-mode theory}

By the temporal coupled-mode theory 
\cite{Joannopoulos2008bk,Manolatou1999}
the resonance is described by the cyclic eigenfrequency $\omega_0$ and complex amplitude $A$. This theory should not be confused with the spatial coupled-wave theory where the amplitudes of propagating waves are involved
\cite{Kogelnik1969,Pierce1954}.
The resonance manifests itself through the amplitudes $s_{\ell\pm}$ of incoming and outgoing energy fluxes:
	\begin{eqnarray}
	\frac{dA}{dt}=
	-i\omega_0 A
	-\sum_{\ell=1}^N \frac{A}{\tau_\ell}
	+\sum_{\ell=1}^N \sqrt{\frac{2}{\tau_\ell}} s_{\ell +}, \nonumber \\
	s_{\ell -} = - s_{\ell +} + \sqrt{\frac{2}{\tau_\ell}} A.
	\label{eq:CMT}
	\end{eqnarray}
Excitation via one of the ports $s_{\ell+}=s_0 \exp{(-i\omega t)}$ yields the amplitude
	\begin{equation}
	A_\ell(\omega) = \frac{\sqrt{\frac{2}{\tau_\ell}} }
	{i(\omega-\omega_0) +
	\sum_{\ell=1}^N {\frac{1}{\tau_\ell}}} s_{\ell+}.
	\label{eq:Amplitude}
	\end{equation}
Amplitudes of reflection from port $\ell$ into port $\ell'$ form a scattering matrix
	\begin{equation}
	r_{\ell\ell'} = \frac{s_{\ell'-}}{s_{\ell+}} = 
	-\hat{\delta}_{\ell\ell'} + \frac{\sqrt{\frac{2}{\tau_\ell}}\sqrt{\frac{2}{\tau_\ell'}}}
	{i(\omega-\omega_0) +
	\sum_{\ell''=1}^N {\frac{1}{\tau_\ell''}}},
	\label{eq:SM}
	\end{equation}
where $\hat{\delta}_{\ell\ell'}$ is the Kronecker symbol. Reflections are observed as spectral peaks in the shape of Lorentz profiles with full width at half maximum <FWHM>
	\begin{equation}
	2\gamma = 2\sum_{\ell=1}^N {\frac{1}{\tau_\ell}}.
	\label{eq:FWHM_omega}
	\end{equation}
Scaling from the cyclic frequency $\omega = 2\pi\nu$ to the frequency $\nu$ yields a $2\pi$ times narrower width of the peak (\ref{eq:FWHM_omega}):
	\begin{equation}
	\Delta\nu =  \frac{\gamma}{\pi} = \frac{1}{\pi}\sum_{\ell=1}^N {\frac{1}{\tau_\ell}}.
	\label{eq:FWHM}
	\end{equation}
This is sufficient to allow the spectral behavior of the state to be described in terms of temporal coupled-mode theory.
%
% \begin{widetext}
\begin{figure*}[htbp]%[htbp]
\includegraphics[scale=.99]{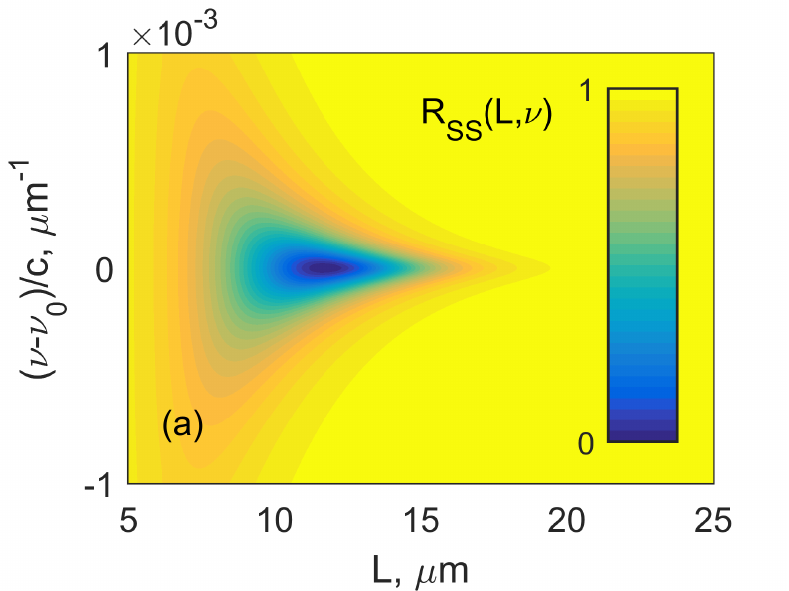}
\includegraphics[scale=.99]{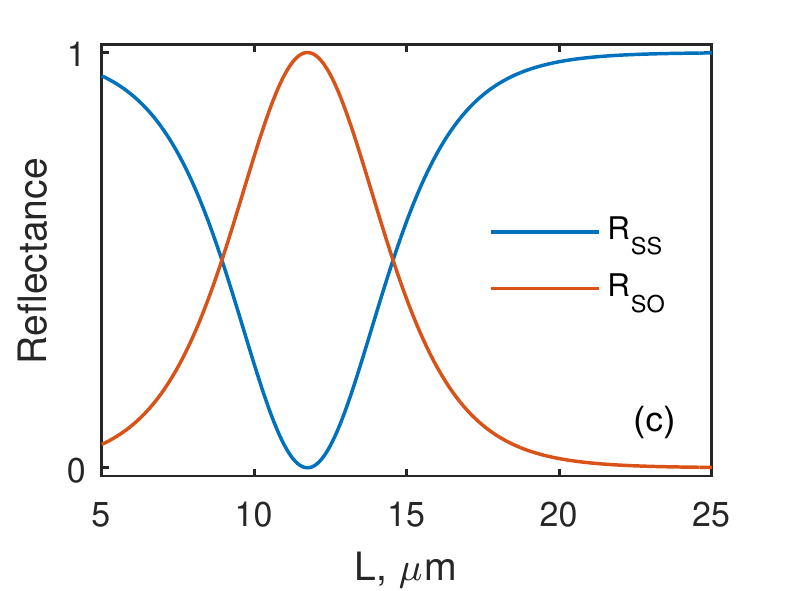}
\includegraphics[scale=.99]{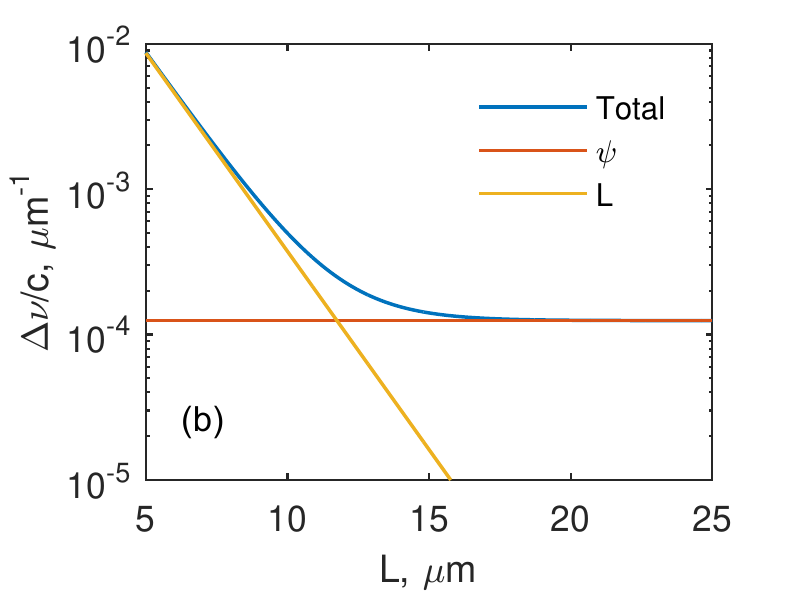}
\includegraphics[scale=.99]{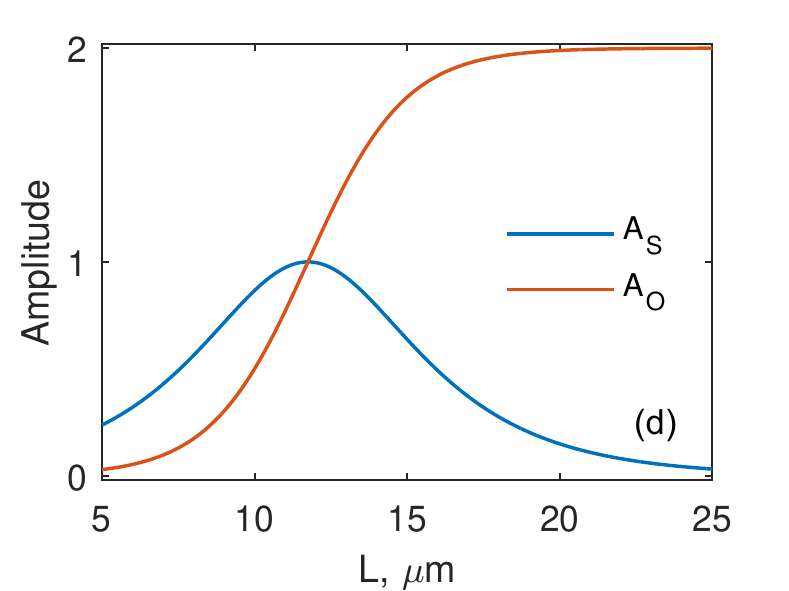}
\caption{%
The Kopp--Genack effect. 
The linewidth saturates with increasing cholesteric layer thickness. 
Polarization reversal of the optimal exciting light. 
(\textbf{a}) Reflection spectrum, Eq. 
(\ref{eq:SM}). 
(\textbf{b}) Spectral dip width, Eq. 
(\ref{eq:FWHM}). 
The cholesteric helix pitch is $p = 1\mu m$, electric anisotropy is $\delta_\epsilon = 0.1$; there is no magnetic anisotropy, $\delta_\mu = 0$. 
% The chart scale is not adequate to show discrepancy in the results obtained by the TCMT and Berreman methods
(\textbf{c}) Reflection at the resonance frequency $\omega = \omega_0$, Eq. 
(\ref{eq:SM_max}). 
(\textbf{d}) State amplitude at the resonance frequency $\omega = \omega_0$, Eq. 
(\ref{eq:Amplitude}). 
% The parameters are the same as above in Fig.~\ref{Fig3}. 
% The chart scale is not adequate to show discrepancy in the results obtained by the TCMT and Berreman methods
}
\label{Fig3}
\end{figure*}
% \end{widetext}

In the previous section we dwelt in detail on ideal COTS with its eigenfrequency given by (\ref{eq:angle_dispersion}). 
Relaxation times (\ref{eq:t_psi}) and (\ref{eq:t_L}) were used to express deviations from the ideal state associated, respectively, with the angle $\psi$ and the layer thickness $L$. 
This appears to be sufficient to fully describe COTS in terms of TCMT. 
The port of coupling via angle $\psi$ has the cross-handed circular polarization with respect to the cholesteric twist while coupling via the cholesteric thickness $L$ has the co-handed circular polarization. 
Equation (\ref{eq:SM}) at the resonance frequency $\omega = \omega_0$ generates the following reflection matrix:
	\begin{equation}
	\hat{R} = 
	\left[\begin{array} {cc}
	r_{SS}^2 		& r_{SO}^2 \\ 
	r_{OS}^2 		& r_{OO}^2 \\ 
	\end{array} \right] =
	\left[\begin{array} {cc}
	\cos^2 2\theta		& \sin^2 2\theta \\ 
	\sin^2 2\theta		& \cos^2 2\theta \\ 
	\end{array} \right],
	\label{eq:SM_max}
	\end{equation}
where $\tan \theta = \tau_\psi / \tau_L$, indices $S$ and $O$ stand for the same and the opposite circular polarizations, respectively. 
This matrix satisfies the energy conservation law. 
Maximum nondiagonal reflections occur at crossover when the relaxation times become equal $\theta=\pi/4$. 

An analytical expression for the length of crossover $L_c$ is obtained by equalizing relaxation times (\ref{eq:t_psi}) and (\ref{eq:t_L}), which yields:
	\begin{equation}
	\exp(
	\frac{4\pi L_c}{\lambda_f}
	) =
	\frac{1}{\sin^2 (\psi/2)},
	\end{equation}
	\begin{equation}
	L_c = 
	\frac{\lambda_f}{4\pi}
	% \frac{\lambda_0}{2\pi \delta_\epsilon \sin 2\bar{\chi}}
	|\log{\sin^2 (\psi/2)}|. 
	\label{eq:L_crossover}
	\end{equation}
This length, as is fairly noted in \cite{Kopp2003_comment}, is difficult to measure in the twist defect of a cholesteric because of the high $\tau_\psi$ and the high requirements to the experiment accuracy. 
In the case of COTS, the $\tau_\psi$ magnitude appears to be substantially larger because of the HPM imperfectness and the crossover length reduces, which should simplify its experimental measuring. 
\subsection*{TCMT applicability limits}
When using TCMT to tackle the problem of coupling between a localized state and waveguides, there are certain limits imposed such as linearity, time-invariant structural parameters, energy conservation, time-reversible energy flux and weak coupling \cite{Joannopoulos2008bk}. 
The model chosen a priori meets the first four requirements. 
The fifth one incorporates, in fact, two requirements. 
First, the solution derived from (\ref{eq:CMT}) ignores corrections with respect to a small parameter that is defined as a vibration period-to-relaxation time ratio \cite{Haus1984bk}. 
Second, the waveguide dispersion should not be large over the frequency range of the resonance Lorentz profile. 
For a fairly large relaxation time the Lorentz profile width tends to zero and this automatically takes care of the second requirement \cite{Joannopoulos2008bk}. 
In this case, the cholesteric acts as a waveguide for waves with cross-handed circular polarization. 
Dispersion in the middle of the stop band is moderate. 
So, we believe the set of linear differential equations (\ref{eq:CMT}) provides an adequate description of the model when $L\ge 5p$ and $\delta_\epsilon=0.1$. 
\section*{Results}
Fig.~\ref{Fig3} shows that there is a Kopp--Genack effect observed for COTS. 
For definiteness, we took the helix pitch to be  $p=1 \mu\mbox{m}$. 
For the parameters specified in the figure, the eigenfrequency $\nu_0/c=1\mu\mbox{m}^{-1}$ is achieved at the angle $\phi = \bar{\chi} \approx 45.7^\circ$, the mismatch angle being $\psi \approx 2.9^\circ$. 
The crossover length $L_c \approx 11.75 \mu\mbox{m}$ agrees with Eq. 
(\ref{eq:L_crossover}). 
For substantially thicker cholesteric layers, $L\gg L_c$, the line width is $\Delta\nu_\infty/c \approx 1.25\cdot 10^{-4} \mu\mbox{m}^{-1}$ and the quality factor saturates $Q_\infty=\nu_0 /\Delta\nu_\infty \approx 8000$. 
Note that for crossover, $L = L_c$, the quality is half that magnitude. 
Fig.~\ref{Fig3}a illustrates that for crossover $L = L_c$, the structure reflects light as a conventional mirror, i.e. 
the circular polarization handedness changes for the opposite one. 
The growing thickness of the cholesteric layer $L\gg L_c$ takes its polarization properties back to the HPM properties when $L=0$, i.e. 
the circular polarization handedness no longer changes under reflection. 
Fig.~\ref{Fig3}b proves that the co-handed and cross-handed (right and left) circular polarizations excite COTS equally efficiently when $L = L_c$ (crossover). 
Hence, the most efficient excitation occurs when both polarizations with equal amplitudes and a certain phase difference are superimposed to produce linearly polarized light. 
No COTS is excited by orthogonally polarized light. 
For thicker cholesterics, $L\gg L_c$, excitation is only possible with cross-handed circularly polarized light as co-handed circularly polarized light is totally reflected from the cholesteric.

\section*{Conclusion}
The chiral optical Tamm state can be considered strictly localized only provided the electric and magnetic permittivities are equal and the respective tensor axes coincide. 
Otherwise COTS is observed as a polarized reflection resonance with two relaxation constants determined by the permittivity difference and the cholesteric layer thickness. 
COTS is closely connected with the twist defect of cholesteric and renders the Kopp--Genack crossover effect. 
A scattering matrix, where spectral peaks are described by Lorentz profiles, has been found in terms of the temporal coupled-mode theory. 
It is for the first time that a formula for the cholesteric crossover thickness has been suggested, which is equally suitable for COTS and for the twist defect. 
The analytical result agrees with the direct numerical one. 

This work was financially sponsored by the Ministry of Science and Technology (MOST), 
 Taiwan, under grant No. <coming soon>, 
Russian Foundation for Basic Research, Government of Krasnoyarsk Territory, 
Krasnoyarsk Region Science and Technology Support Fund to the research project No. 17-42-240464, 
 and by the Siberian Branch of the Russian Academy of Sciences 
 under Complex Program II.2P (projects Nos. 0356-2015-0410 and 0356-2015-0411). 
The authors are thankful to 
Prof. A.F.~Sadreev, Prof. E.N.~Bulgakov, 
Dr. N.V.~Rudakova,
Y.-C.~Hsiao and L.V.~Pertseva for valuable discussions and comments.

\bibliography{library}
\end{document}

\title{Chiral Optical Tamm States: Temporal Coupled-Mode Theory}

\author{Ivan V. Timofeev}
\email{tiv@iph.krasn.ru}
\affiliation{Kirensky Institute of Physics, Siberian Branch of the Russian Academy of Sciences, Krasnoyarsk 660036, Russia}
\affiliation{Laboratory for Nonlinear Optics and Spectroscopy, Siberian Federal University, Krasnoyarsk 660041, Russia}
\author{Pavel S. Pankin}
\affiliation{Institute of Engineering Physics and Radio Electronics, Siberian Federal University, Krasnoyarsk 660041, Russia}
\author{Stepan Ya. Vetrov}
\affiliation{Kirensky Institute of Physics, Siberian Branch of the Russian Academy of Sciences, Krasnoyarsk 660036, Russia}
\affiliation{Institute of Engineering Physics and Radio Electronics, Siberian Federal University, Krasnoyarsk 660041, Russia}
\author{Vasily G. Arkhipkin}
\affiliation{Kirensky Institute of Physics, Siberian Branch of the Russian Academy of Sciences, Krasnoyarsk 660036, Russia}
\affiliation{Laboratory for Nonlinear Optics and Spectroscopy, Siberian Federal University, Krasnoyarsk 660041, Russia}
\author{Wei Lee}
\affiliation{Institute of Imaging and Biomedical Photonics, College of Photonics, National Chiao Tung University, Guiren Dist., Tainan 71150, Taiwan}
\author{Victor Ya. Zyryanov}
\affiliation{Kirensky Institute of Physics, Siberian Branch of the Russian Academy of Sciences, Krasnoyarsk 660036, Russia}

\date{\today}

\begin{abstract}
Хиральное оптическое таммовское состояние (ХОТС)
представляет собой локализованное состояние на границе 
сохраняющего хиральность зеркала 
и структурно хиральной среды,
% SCM - structurally chiral medium
такой как холестерический жидкий кристалл или
хиральная скульптурированная тонкая пленка.
% Chiral STF - Sculptured thin film
Спектральное проявление ХОТС
в виде резонансов отражения
описывается при помощи теории связанных мод.
Для излучений двух круговых поляризаций связь с модой различна,
поскольку ХОТС имеет геликоидальную структуру, повторяющую структуру холестерика.
% co-handed circularly polarized light over the cross-handed
Для излучения с круговой поляризацией того же знака
связь экспоненциально ослабляется с толщиной холестерического слоя,
поскольку частота ХОТС находится в стоп-зоне.
Излучение с круговой поляризацией противоположного знака
беспрепятственно проходит сквозь холестерический слой,
и может возбудить ХОТС при отражении от сохраняющего хиральность зеркала;
связь пропорциональна анизотропии холестерика и
теоретически может быть полностью прекращена
только при наличии анизотропии магнитной проницаемости.
Равенство этих двух связей приводит к поляризационному кроссоверу (эффект Коппа-Генака
\hide{Kopp–Genack effect}),
при котором для возбуждения ХОТС оптимальна линейная поляризация.
Соответствующая кроссоверу толщина холестерика
и матрица рассеяния для ХОТС описываются в общем виде простыми выражениями.
%Спектральные пики аппроксимируются лоренцевыми контурами.
\end{abstract}

% \pacs{42.70.Df, 61.30.Gd, 42.79.Ci, 42.60.Da, 42.87.Bg}

\maketitle

\section*{Введение}
\begin{figure}[b]%[htbp]
\center{\includegraphics[scale=.75]{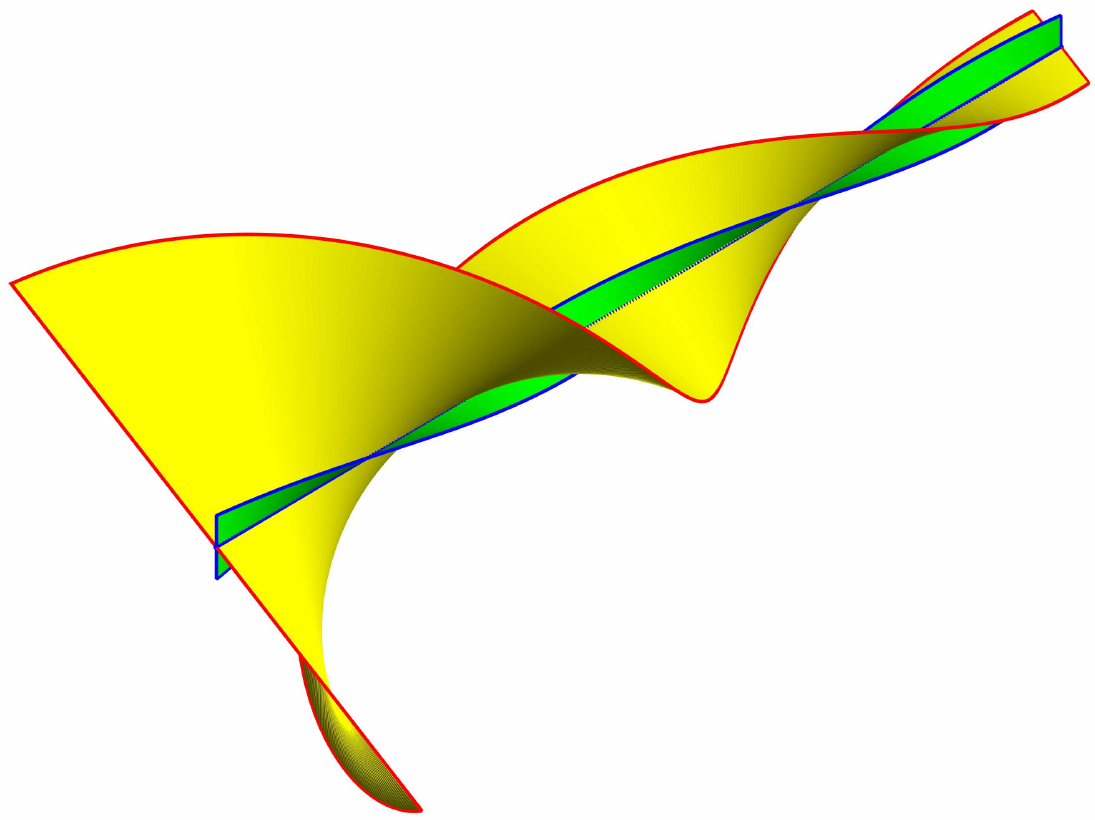}}
\caption{%
Круговая дифракция Брэгга формирует ХОТС.
Директор холестерика изображен синим и зеленым,
напряженность электрического поля изображена красным и желтым.
Угол \hide{$\chi$} между ними не меняется с глубиной.}
% Sketch of a chiral OTS, electric field drawn in red-yellow, cholesteric director drawn in dlue-green.
% Angle $\phi$ between them does not depend on depth.}
\label{Fig:OTS}
\end{figure}

Материя склонна упорядочиваться, образуя кристаллы. 
В жидких кристаллах первенствует ориентационный порядок,
благодаря цикличности способный порождать эхо трансляционного порядка
в хиральных сверхрешетках холестерического жидкого кристалла.
При дифракции в холестериках
поле запечатлевает винтовой порядок материи,
поскольку дифрагирует волна с сонаправленной круговой поляризацией (Рис.~\ref{Fig:OTS}),
а волна противоположно направленной круговой проляризации 
распространяется без заметных изменений.
Это явление называют селективным отражением
\cite{Belyakov1992b},
либо круговой дифракцией Брэгга
для электромагнитных и акустических волн \cite{Faryad_Lakhtakia2014rv}.
Представленные здесь результаты естественно обобщаются
на любые материалы с геликоидальным откликом,
включая широко перестраиваемые геликонические структуры
\cite{Xiang_Lavrentovich2015}.
% структурно хиральной среды,
% SCM - structurally chiral medium
% Chiral STF - Sculptured thin film
% геликонус
Полупрозрачность запрещенной зоны затрудняет
наблюдение локализованных состояний 
при нарушении порядка на границе или дефекте структуры.
Взамен, упрощается описание состояний,
из-за преимущества гладкости геликоидальной симметрии
\cite{Belyakov1992b,Avendano_Oldano2005}
в сравнении с дискретностью трансляционной симметрии кристаллов.
Дефект структуры принято упрощенно представлять как
полость, ограниченную зеркалами,
роль которых выполняет решетка Брэгга.
Такой аналог резонатора Фабри-Перо обладает рядом дефектных мод -- локализованных оптических состояний,
обычно соответствующих целому числу полуволн, умещающихся в полости.
Анизотропия холестерика и дефекта приводит к ряду поляризационных особенностей
\cite{Haus1976,Lakhtakia1999_Defect,Yang1999,Shabanov2004,Gevorgyan2006,Belyakov2011,Kiselev2014},
связанных с приложениями
\cite{Faryad_Lakhtakia2014rv,Hsiao2014,Rodarte2015,LinJiaDe2015,Kesaev2017},
среди которых выделяется дефект кручения,
который не содержит промежуточного слоя и имеет нулевую толщину 
\hide{spaceless, spacerless twist-defect}
\cite{Hodgkinson2000_Twist,Kopp2002}. % своеобразие самобытность
Его строгое теоретическое
\cite{Oldano2003,Becchi_Oldano2004,Avendano_Oldano2005,Schmidtke2003_EPJ}
и экспериментальное 
\cite{Hodgkinson2000_Twist,Schmidtke2003_PRL,Ozaki2003_CLC_laser,Shibaev2002_CLC_laser}
исследование
породило дискуссию о поляризации и времени релаксации локализованного состояния
\cite{Oldano2003_comment,Kopp2003_comment}.
Теоретически бесконечное время релаксации 
возможно лишь при наличии анизотропии магнитной проницаемости
\cite{Avendano_Oldano2005,Gevorgyan_Rafayelyan2013}.
При отсутствии таковой 
оказывается невозможным согласовать 
одновременно электрическую и магнитную напряженности на границе.
И бесконечное увеличение толщины холестерика
не ведет к бесконечной добротности.
Насыщение добротности с увеличением толщины холестерика
приводит к смене круговой поляризации выходящего излучения
с сонаправленной на противоположно направленную.
Этот кроссовер поляризации
\cite{Kopp2002}
получил название эффекта Коппа-Генака 
\cite{McCall_Hodgkinson2015bk}.

В отличие от резонанса Фабри-Перо, дефект кручения 
порождает единственное локализованное состояние.
Пространственное распределение поля при этом состоит из двух волн,
экспоненциально спадающих в противоположных направлениях,
и не содержит плоской вершины \hide{flat-top}.
Это напоминает поверхностную волну, с той лишь разницей, что
на дефекте кручения нет ограничения на угол падения возбуждающей волны,
а поверхностные волны на границе холестерика с изотропным диэлектриком
наблюдаются лишь при углах, обеспечивающих полное внутреннее отражение
\cite{Belyakov1992_Shilina}.
Известна поверхностная волна, не ограниченная полным внутренним отражением,
это оптическое таммовское состояние (ОТС)
\cite{Kavokin2005,Vinogradov2010rv,Iorsh2012,Afinogenov2013,Auguie_Fainstein2015,ChenKuoping2015,ChenKuoping2016}.
ОТС представляет собой аналог таммовского состояния электронов на границе сверхрешетки
и его дисперсия лежит выше светового конуса, описываемого прямой $k=\omega/c$
% The dispersion of the surface plasmon lies outside the light cone given by $k=\omega/c$
\cite{Kaliteevski2007a}.
Такое состояние может возбуждаться даже перпендикулярно поверхности
и не переносить энергии вдоль нее, что удобно в приложениях.
Естественно встает вопрос о наличии ОТС
на границе холестерика с металлом при перпендикулярном падении света.
Здесь следует учитывать две особенности.
Во-первых, это полупрозрачность холестерика из-за круговой дифракции Брэгга.
Во-вторых, это поляризационная неустойчивость,
вызванная чередованием круговых поляризаций:
отражение от холестерика не меняет знак круговой поляризации,
а отражение от металла -- меняет
\cite{Abdulhalim2006}.
Это чередование подобно светофору,
запрещающему движение через холестерик волне совпадающей круговой поляризации,
и открывающему движение после отражения от металла
\cite{Timofeev2013}.
Компенсация изменения поляризации может достигаться
при помощи дополнительного анизотропного слоя.
Такой слой порождает серию локализованных состояний,
обладающих нетривиальными поляризационными свойствами
\cite{Pyatnov2014m_OL,Pyatnov2016m_JOpt,Pyatnov2017m_LC}.
Различные комбинации с двумя зеркалами широко изучались
и предлагались для практических применений
\cite{Zhuang_Patel1999,Abdulhalim2006}, включая хиральный слой с толщиной менее одного шага винта
\cite{Isaacs_Abdulhalim2014,Hsiao2011,Timofeev2012,Timofeev2016t,Gunyakov2016_JQSRT,HuangKuanChung2016m}.
Плотность энергии таких состояний может сгущаться
по мере приближения к зеркалу,
но они локализуются в объеме дополнительного слоя,
а не на поверхности.
Обойтись без дополнительного слоя возможно,
когда используется сохраняющее хиральность зеркало (СХЗ),
по-английски, handedness-preserving mirror (HPM, handed mirror)
\cite{Plum_Zheludev2015,Fedotov_Zheludev2005,Kopp2003_comment,Rudakova2017tm_BRAS}.
Такое зеркало можно назвать иначе, как
сохраняющее поляризацию анизотропное зеркало (СПАЗ)
\cite{Rudakova2017tm_BRAS}.
Иначе можно определить СПАЗ как полуволновую фазовую пластинку, действующую на отражение \cite{Kildishev2015}.
% поскольку
При отражении оно сохраняет не только знак хиральности, но и величину эллиптичности.
% Другими словами, сохраняется состояние поляризации волны целиком.
% К тому же СПАЗ по-русски произносится легче, чем СХЗ.
Локализованное на границе СПАЗ и холестерика состояние
было найдено в приближении малой анизотропии холестерика
и названо хиральным оптическим таммовским состоянием (ХОТС)
\cite{Timofeev2016t}.

В данной работе мы решаем вопрос о принципиальной возможности такого состояния
при наличии идеального СПАЗ и полубесконечного непоглощающего слоя холестерика,
обладающего конечной анизотропией.
Проведено подробное описание простейшего случая,
когда электрическая и магнитная анизотропии равны.
Рассматриваются два типа отклонений, именно, отсутствие магнитной анизотропии
и конечная толщина слоя холестерика.
При этом идеальное состояние становится резонансом
с конечной добротностью и временем релаксации.
\hide{Общность состояний, локализованных на границах и в дефектах геликоидальных структур.}

\subsection*{Метод описания спектральных пиков}

Для расчета спектров и распределения поля
удобен формализованный метод Берремана \cite{Berreman1972,Palto2001}.
При нормальном падении на холестерик есть несложное точное аналитическое решение
\cite{Oseen1933,DeVries1951,Kats1971,Nityananda1973}.
Сшивка тангенциальных напряженностей на границе холестерика
позволяет записать общие замкнутые формулы 
\hide{closed-form equations}
\cite{Oldano2003,Becchi_Oldano2004,Avendano_Oldano2005,Schmidtke2003_EPJ}.
Для большей простоты и ясности физического явления
мы дополнительно используем приближенный аналитический метод, это
временная теория связанных мод (ВТСМ),
или теория связанных мод во временной области
\cite{Haus1984bk,Manolatou2002bk,Joannopoulos2008bk}.
ВТСМ позволяет описывать поле в резонаторах,
связанных с волноводами 
\hide{Coupled resonators; Resonator-waveguide coupling}.
Пространственная структура локализованной моды здесь не рассматривается.
Выписывается комплексная амплитуда этой моды
и ее производная по времени.
Теория широко применяется для стационарных процессов,
в которых производная по времени зануляется.
Первое слово в термине, ``временная''
\cite{FanShanhui2003},
можно рассматривать как исторически сложившееся.
По сути тот же метод используется для описания открытых резонаторов
\cite{Xu_Yariv2000,Bliokh2008rv},
и восходит к решению Липпманом и Швингером квантовомеханической задачи рассеяния
\cite{Lippmann1950}.
Этот метод следует отличать от 
теории связанных волн
\cite{Kogelnik1969,Yariv1973_CMT},
или пространственной теории связанных мод,
широко используемой в оптике холестериков
\cite{Belyakov1974,Belyakov1979,Belyakov1992_Shilina,Belyakov1992t_ShilinaScalar,McCall2000,Wang2005}.
Уравнения последней теории описывают связанные амплитуды распространяющихся волн,
и здесь не идет речи об амплитуде моды резонатора.
Хотя в основе обоих подходов лежит понятие связанных мод \cite{Pierce1954}.

\section*{Модель}

Рисунок~\ref{Fig:OTS} изображает границу холестерика.
Ось закручивания холестерика перпендикулярна поверхности зеркала.
Другими словами,
директор холестерика, единичный вектор преимущественного направления молекул,
постоянен в сечениях, параллельных границе,
и равномерно поворачивается по мере удаления от границы.
Вблизи границы возможно ХОТС, описанное в
\cite{Timofeev2016t} в пределе малой анизотропии холестерика.
Это состояние можно представить как наложение двух волн одинаковой круговой поляризации.
Эти волны бегут навстречу друг другу,
а их напряженности вращаются во времени в противоположных направлениях. 
Результирующая поляризация линейна в каждой точке пространства,
а плоскость поляризации по мере удаления от границы равномерно поворачивается
вместе с директором холестерика.
При этом амплитуда экспоненциально спадает.
Структура поля напоминает
скрученный луковичных купол собора Василия Блаженного
или купол Копенгагенской фондовой биржи.
% It resembles a screwed onion dome of Saint Basil's Cathedral in Moscow or Dutch Renaissance style in Copenhagen Stock Exchange dome.

\begin{figure}[b]%[htbp]
\center{\includegraphics[scale=1]{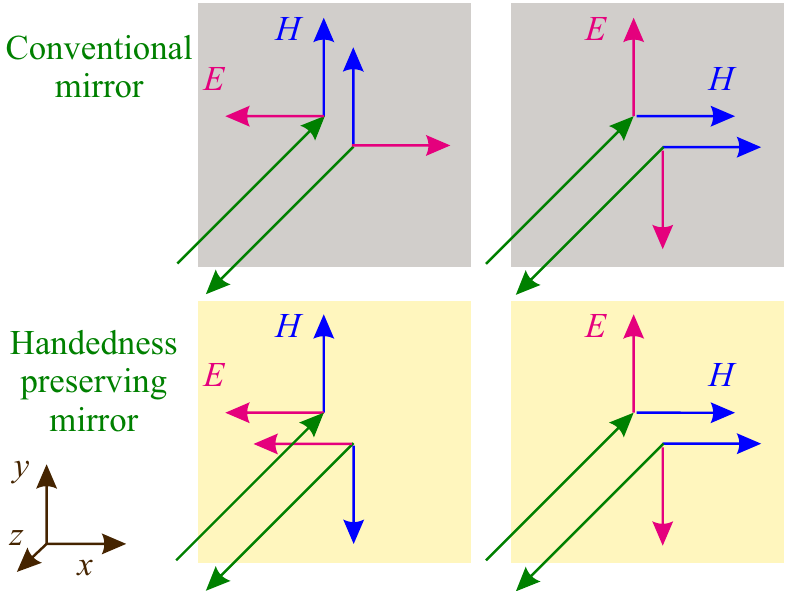}}
\caption{%
СПАЗ и обычное зеркало.
Отражение $x$-поляризованного света различно.
Для СПАЗ электрическая напряженность $E_x$ отраженного поля сохраняет фазу,
вместо этого фазу меняет мангитная напряженность $H_y$.
}
% Sketch of a chiral OTS, electric field drawn in red-yellow, cholesteric director drawn in dlue-green.
% Angle $\phi$ between them does not depend on depth.}
\label{Fig:HPM}
\end{figure}
Рисунок~\ref{Fig:HPM} сопоставляет СПАЗ и обычное зеркало.
В обычном \hide{conventional} зеркале электрическое поле при отражении меняет фазу на противоположную,
а магнитное поле не меняет фазу.
Возможна обратная ситуация, тогда зеркало называют магнитным \hide{magnetic wall}.
СПАЗ сочетает электрический и магнитный тип отражения \cite{Fedotov_Zheludev2005}.
Обозначим через $y$ такую ось в плоскости СПАЗ,
что линейно поляризованное вдоль нее электрическое поле отражается в противофазе.
Ортогональная ось $x$ соответствует магнитному отражению.
Важно, что магнитное поле при этом ортогонально электрическому,
и, следовательно, ортогонально оси $x$, направлено вдоль $y$.
Другими словами, компоненты электрического и магнитного поля, направленные вдоль $y$,
испытывают скачек фазы, и на границе дают узел, нулевую напряженность.
Ненулевыми остаются только $x$-компоненты поля.

\subsection*{Уравнения Максвелла в базисе,\\ связанном с директором холестерика}

Ограничимся случаем нормального падения света.
Уравнение Максвелла для волны, распространяющейся вдоль оси $z$ геликоида, при $\mu=1$
можно записать в виде
	\begin{equation}
	\frac{\partial^2 \vec{E}}{\partial z^2} = 
	\frac{\hat{\epsilon}}{c^2}
	\frac{\partial^2 \vec{E}}{\partial t^2}. 
	\end{equation}
Волна описывается вектором $\vec{E}$ комплексных амплитуд
для компонент электрического поля в ортогональных направлениях $x$ и $y$.
Проекция $\hat{\epsilon}$ тензора диэлектрической проницаемости на плоскость $x,y$ на глубине холестерика $z$ имеет вид
	\begin{equation}
	\hat{\epsilon} = \left[\begin{array} {cc}
	\epsilon_e \cos^2 \tilde{\phi} + \epsilon_o \sin^2 \tilde{\phi} & \sin 2\tilde{\phi} \;  (\epsilon_e-\epsilon_o)/2 \\ 
	\sin 2\tilde{\phi} \; (\epsilon_e-\epsilon_o)/2 & \epsilon_e \sin^2 \tilde{\phi} + \epsilon_o \cos^2 \tilde{\phi} 
	% \epsilon_o \cos^2 \tilde{\phi} + \epsilon_e \sin^2 \tilde{\phi} & \sin 2\tilde{\phi} \;  (\epsilon_e-\epsilon_o)/2 \\ 
	% \sin 2\tilde{\phi} \; (\epsilon_e-\epsilon_o)/2 & \epsilon_o \sin^2 \tilde{\phi} + \epsilon_e \cos^2 \tilde{\phi} 
	\end{array} \right]. 
	\end{equation}
Здесь оптическая ось, совпадающая с директором холестерика, описывается углом закручивания $\tilde{\phi}(z) = 2\pi z /p +\phi$, который отсчитывается от оси $x$ в направлении оси $y$;
$p$ – шаг геликоида \hide{винтовой спирали}. Положительным значениям шага соответствует правозакрученный геликоид,
а отрицательным -- левозакрученный.

Учет магнитной проницаемости $\hat{\mu}_{xy}$ 
делает предпочтительной явную запись магнитной напряженности $\vec{H}$
и увеличивает размерность вектора поля с 2 до 4.
	\[
	\vec{J} = [E_x,H_y,E_y,-H_x]^T.
	\]
Соответственно, порядок дифференциального уравнения понижается со второго до первого.
Предположим, что главные оси тензоров магнитной и электрической проницаемостей совпадают.
Это позволяет перейти в ортонормированный базис $u,v,z$,
равномерно вращающийся вместе с директором холестерика,
так что ось $u$ всегда направлена вдоль директора:
	\[
	\vec{J}_R = [E_u,H_v,E_v,-H_u]^T.
	\]
Согласно методу Берремана \cite{Berreman1972} в стационарном случае уравнения Максвелла принимают форму:
	\begin{equation}
	\frac{\partial \vec{J}_R}{\partial \tilde{z}} = 
	i \hat{T}\vec{J}_R. 
	\label{eq:Berreman}
	\end{equation}
Дифференциальная матрица переноса \hide{differential propagation matrix}
для вращающегося басиза выписана в \cite{Oldano2003,Avendano_Oldano2005} 
и может быть приведена к виду:
	\begin{equation}
	\hat{T} = \left[\begin{array} {cccc}
	0 & \mu_o & \tilde{\lambda} & 0 \\ 
	\epsilon_e & 0 & 0 & \tilde{\lambda} \\ 
	\tilde{\lambda} & 0 & 0 & \mu_e \\ 
	0 & \tilde{\lambda} & \epsilon_o & 0 \\ 
	\end{array} \right],
	\label{eq:TM}
	\end{equation}
где $\tilde{\lambda} = \lambda_0/p =  2 \pi c / \omega p$ -- обезразмеренная длина волны, $\tilde{z} = z \omega/c=2\pi z/ \lambda_0$ -- обезразмеренная координата.
В системе СИ размерности электрической и магнитной напряженностей не равны и должны нормироваться через импеданс вакуума $Z_0=E/H=\sqrt{\mu_0/\epsilon_0}$.

\subsection*{Решение без приближения малой анизотропии}

Уравнению (\ref{eq:Berreman}) соответствуют четыре нормальные волны.
Они определяются собственными числами матрицы $\hat{T}$,
имеющими смысл показателей преломления $n$.
Соответствующие собственные векторы матрицы $\hat{T}$
имеют смысл поляризаций $\vec{J_0}$.
Симметрия относительно обращения направления оси геликоида
позволяет классифицировать эти четыре нормальные волны как 
две пары противоположно направленных волн.
В каждой паре меньшую фазовую скорость имеет волна с б\'{о}ольшим показателем преломления,
будем называть ее медленной (slow). А другую волну -- быстрой (fast): 
	\begin{equation}
	\vec{J}_{R0} = \vec{J}^\pm_{s,f}\exp(\pm i n_{s,f}\tilde{z}). 
	\label{eq:J_eigen}
	\end{equation}
Подстановка решений $\vec{J}^+_{s,f}$ в уравнение (\ref{eq:Berreman}) дает 
	\begin{equation}
	i [\hat{T} - n_{s,f} \hat{I}] \vec{J}_{s,f} = 0,
	\label{eq:eigen}
	\end{equation}
где $\hat{I}$ -- единичная матрица, индекс `$+$' в $\vec{J}^+_{s,f}$ опущен.
И показатели преломления
	\begin{equation}
	n_{s,f}^2 = \tilde{\lambda}^2 + \overline{\epsilon\mu}
	\pm \sqrt{4\bar{\epsilon}\bar{\mu}\tilde{\lambda}^2+d_{\epsilon\mu}^2},
	\label{eq:eigenvalue}
	\end{equation}
где $d_{\epsilon\mu}=(\epsilon_e\mu_o-\epsilon_o\mu_e)/2$ -- \emph{анти}симметрия проницаемостей,
а верхняя черта означает среднее арифметическое по обыкновенной и необыкновенной проницаемостям:
	\begin{equation}
	\bar{\epsilon}=\frac{\epsilon_e+\epsilon_o}{2},
	\bar{\mu} = \frac{\mu_o+\mu_e}{2},
	\overline{\epsilon\mu}=\overline{\epsilon_{eo}\mu_{oe}}=\frac{\epsilon_e\mu_o+\epsilon_o\mu_e}{2}. 
	% \label{eq:eigenvalue}
	\end{equation}

Использование масштабной инвариантности \hide{scale invariance} уравнений Максвелла (\ref{eq:Berreman})
и нормировка материальных параметров
(см. Приложение в \cite{Avendano_Oldano2005})
сводит структуру к двум определяющим параметрам.
Это анизотропии электрической и магнитной проницаемостей
	\begin{equation}
	\delta_\epsilon = \frac{\epsilon_e-\epsilon_o}{\epsilon_e+\epsilon_o},
	\delta_\mu = \frac{\mu_e-\mu_o}{\mu_e+\mu_o}.
	\end{equation}
\subsection*{Случай равных анизотропий, $\delta_\epsilon = \delta_\mu$}

Рассмотрим случай равных анизотропий $\delta = \delta_\epsilon=\delta_\mu$.
Ясность и изящество этого однопараметрического семейства структур
должны компенсировать в глазах читателя трудности его физической реализации в оптическом диапазоне
\cite{Gevorgyan_Rafayelyan2013}.
Длина волны параметризует семейство дифференциальных матриц переноса (\ref{eq:TM}) подобно тому,
как эксцентриситет параметризует семейство конических сечений.
\hide{В качестве аналога эллипсов, парабол и гипербол можно рассмотреть три примера $\tilde{\lambda}=0,1,2$
для проницаемостей $\epsilon_o=\mu_e=1, \epsilon_e=\mu_o=3$. Добавить рис. 3 с отражениями от обоих осей}

Дисперсионное уравнение упрощается за счет симметрии проницаемостей $d_{\epsilon\mu}=0$:
	\begin{equation}
	n_{s,f}^2 = (\tilde{\lambda} \pm \sqrt{\bar{\epsilon}\bar{\mu}})^2-(\bar{\epsilon}\bar{\mu}-\overline{\epsilon\mu}),
	% \label{eq:eigenvalue_symmetric}
	\end{equation}
Без дальнейшей потери общности принимаем нормировку
$\bar{\mu} = \bar{\epsilon}$,
$\overline{\epsilon\mu} = 1$.
Тогда
		$\sqrt{\bar{\epsilon}\bar{\mu}} = \bar{\epsilon} \ge 1$.
Другими словами, проницаемость нормируется на среднее геометрическое
$\epsilon_o$ и $\epsilon_e$: $\bar{\epsilon}_g = \sqrt{\epsilon_o \epsilon_e}=1$,
при этом их среднее арифметическое оказывается не меньше единицы.
Второе слагаемое в правой части дисперсионного уравнения %(\ref{eq:eigenvalue_symmetric})
становится квадратом анизотропии $\bar{\epsilon}\bar{\mu}-\overline{\epsilon\mu} = \delta^2$.
Анизотропия получает смысл среднеквадратичного отклонения проницаемостей:
	\begin{equation}
	n_{s,f}^2 = (\tilde{\lambda} \pm \bar{\epsilon})^2-\delta^2. 
	\label{eq:eigenvalue_symmetric}
	\end{equation}
% В работе [Шилина Г.И. 1992 диссертация к.ф.-м.н. на стр 19]
% этот случай назван двухволновым приближением. 
Дисперсионные кривые представляют собой гиперболы,
а в запрещенной зоне -- круг, в отличие от типичного для периодических сред приближения параболой.
Дисперсионное уравнение (\ref{eq:eigenvalue_symmetric}) можно записать
как для показателя преломления, так и для волнового вектора
	\begin{eqnarray}
	n_{s,f}^2 = \tilde{\lambda}^2 \pm 2\bar{\epsilon}\tilde{\lambda} + 1,
	\nonumber \\
	\tilde{k}_{s,f}^2 = n_{s,f}^2/\tilde{\lambda}^2 = \tilde{\omega}^2 \pm 2\bar{\epsilon}\tilde{\omega} + 1. 
	\label{eq:eigenvalue_symmetric_wk}
	\end{eqnarray}

\begin{figure}[htbp]
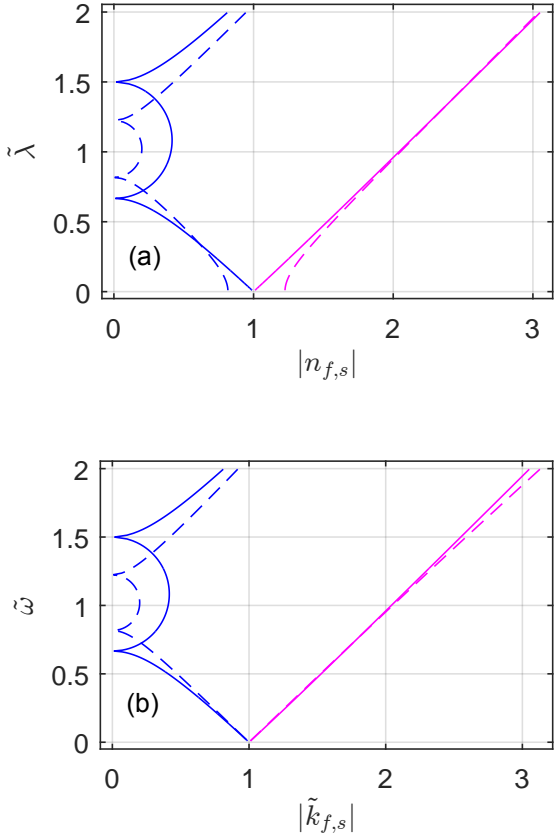

\includegraphics[scale=1]{F3_lambda_n.pdf}
\includegraphics[scale=1]{F3_omega_k.pdf}
\caption{%
Симметрия дисперсионных кривых. 
(a) Длина волны $\tilde{\lambda}$ как функция показателя преломления $|n_{f,s}|$.
(b) Частота $\tilde{\omega} = 1 / \tilde{\lambda}$ как функция волнового числа $|\tilde{k}_{f,s}|$.
Синий цвет -- быстрая волна, пурпурный -- медленная.
Сплошная линия -- $\epsilon_e=\mu_e=3/2, \epsilon_o=\mu_o=2/3$, $\delta_\epsilon=\delta_\mu$.
Полуокружность - дифрагирующая волна,
для которой показатель преломления принимает чисто мнимые значения
$|n_{f}|=Im(n_{f})$.
Режим Могена $\tilde{\lambda} \ll 1$ эквивалентен режиму гомогенизации $\omega \ll 2\pi$ 
в соответствии с дисперсионными уравнениями (\ref{eq:eigenvalue_symmetric_wk}).
Штриховая линия -- $\epsilon_o=2/3, \epsilon_e=3/2, \mu_e=\mu_o=1$. Симметрия нарушается.
}
\label{Fig:BandStructure}
\end{figure}

Эта симметрия дисперсий $\tilde{\lambda}(n)$ и $\tilde{\omega}(\tilde{k}) = 1 / \tilde{\lambda}$
говорит о симметрии длинноволнового и коротковолнового пределов.
В длинноволновом пределе наступает гомогенизация среды и проявление анизотропии падает.
Отрицательная оптическая активность исчезает с приближением к случаю статического поля.
В коротковолновом пределе волноводный режим Могена обеспечивает симметричную положительную оптическую активность
(Рис.~\ref{Fig:BandStructure}).
Нарушение условия равных анизотропий $\delta_\epsilon \neq \delta_\mu$ разрушает эту симметрию.
Обыкновенная и необыкновенная волны становятся различимыми
в высокочастотном пределе, эквивалентном раскручиванию геликоида.

Сосредоточимся на круговой дифракции Брэгга,
когда $(\tilde{\lambda} - \sqrt{\bar{\epsilon}\bar{\mu}})^2 \le \delta^2$
и для быстрой волны показатель преломления $n_f$ принимает чисто мнимые значения.
Фазовая скорость при этом обращается в бесконечность, а групповая теряет смысл.
Здесь уместно записать дисперсионное уравнение (\ref{eq:eigenvalue_symmetric}) в виде тригонометрического тождества,
где вместо длины волны выступает некоторый угол $\chi \in [0,\pi/2]$,
физический смысл которого будет раскрыт позднее.
	\begin{equation}
	- \delta^2 \sin^2 2\chi = \delta^2 \cos^2 2\chi - \delta^2. 
	\end{equation}
Это означает, что при длине волны 
	\begin{equation}
	\tilde{\lambda}_0 = \bar{\epsilon}+ \delta \cos 2\chi,
	\label{eq:angle_dispersion_symmetric}
	\end{equation}
показатель преломления быстрой волны $n_f = i \delta \sin 2\chi$
описывает полное отражение в объеме холестерика. 
Такую волну принято называть дифрагирующей
\cite{Belyakov1992b}.
Поляризация этой волны определяется из нетривиальной разрешимости уравнения 
(\ref{eq:eigen}).
Вычитание показателя преломления из матрицы переноса (\ref{eq:TM}) дает тождество:
\begin{widetext}
	\begin{eqnarray*}
	[\hat{T}-n_f \hat{I}] \vec{J}_f \equiv 0 = \\
	\delta \left[\begin{array} {cccc}
	-i \sin 2\chi 		& \Delta-\bar{\epsilon} 		& - i \Delta - i \cos 2\chi 		& 0 \\ 
	\Delta+\bar{\epsilon} 		& -i \sin 2\chi  		& 0 		& - i \Delta - i \cos 2\chi \\ 
	i \Delta + i \cos 2\chi 		& 0 		& -i \sin 2\chi  		& \Delta+\bar{\epsilon} \\ 
	0 		& i \Delta + i \cos 2\chi 		& \Delta-\bar{\epsilon} 		& -i \sin 2\chi  \\ 
	\end{array} \right] 
	\left[\begin{array} {r}
	\cos \chi \\ 
	- i \sin \chi \\ 
	- \sin \chi \\ 
	- i \cos \chi \\ 
	\end{array} \right] ,
	% \label{eq:TM}
	\end{eqnarray*}
\end{widetext}
где $\Delta = \bar{\epsilon}/\delta$.
Проявляется симметрия относительно обмена 
электрических и магнитных полей и проницаемостей, $E_u/E_v=(-H_v/H_u)^*$.
Параллельность электрической и магнитной напряженностей равносильна тому, что
вектор Умова-Пойнтинга обращается в ноль.
Электрическое и магнитное поля линейно поляризованы под одинаковым углом $-\chi$ к директору холестерика.
Это направление совпадает с осью $x$ при $\chi = \phi = \tilde{\phi}(z=0)$.
Условия на поверхности идеального СПАЗ в точности такие же: как электрическое, так и магнитное поле линейно поляризованы вдоль магнитной оси зеркала (Рис.~\ref{Fig:HPM}).
Следовательно, найденное здесь решение точно описывает ХОТС при конечной анизотропии.
\subsection*{Нарушение равенства анизотропий, $\delta_\epsilon \neq \delta_\mu$}

В общем случае обозначим направления электрической и магнитной поляризаций углами $\chi_E$ и $\chi_H$.
Общее нетривиальное решение (\ref{eq:eigen}) дает собственные вектора
матрицы переноса (\ref{eq:TM})
и точные выражения для углов \cite{Becchi_Oldano2004}:
	\begin{eqnarray}
	&&\tan{\chi_E} =
	\frac{E_u}{E_v} =
	- \tilde{\lambda}_0 n_f \frac{\mu_e+\mu_o}{\epsilon_e \mu_e \mu_o - n_f^2 \mu_o - \tilde{\lambda}_0^2 \mu_e},
		\nonumber  \\
	&&-\cot{\chi_H} =
	\frac{H_v}{-H_u} =
	- \tilde{\lambda}_0 n_f \frac{\epsilon_e+\epsilon_o}{\mu_e \epsilon_e \epsilon_o - n_f^2 \epsilon_o - \tilde{\lambda}_0^2 \epsilon_e},
		\nonumber \\
	&&\psi = \chi_E - \chi_H \neq 0,
		\nonumber \\
	&&\bar{\chi} = (\chi_E + \chi_H)/2,
	\end{eqnarray}
где показатель преломления $n_f$ следует подставлять из дисперсионного уравнения (\ref{eq:eigenvalue}).
Видно, что для чисто мнимого показателя преломления $n_f$ поляризации линейны.
А для действительного показателя преломления $n_s$ поляризации в стоп-зоне почти круговые.

Дисперсионное уравнение (\ref{eq:angle_dispersion_symmetric}) можно обобщить в виде
	\begin{equation}
	\tilde{\lambda}_0 \approx \sqrt{\bar{\epsilon}\bar{\mu}}+\sqrt{\bar{\epsilon}\bar{\mu}-\overline{\epsilon\mu}} \cos 2\bar{\chi}. 
	\label{eq:angle_dispersion}
	\end{equation}
Для холестерика без магнитной анизотропии, $\mu=1$:
	\begin{equation}
	\tilde{\lambda}_0 \approx {\sqrt{\bar{\epsilon}}+\sqrt{\bar{\epsilon}-1} \cos 2\bar{\chi}}. 
	\label{eq:angle_dispersion_mu1}
	\end{equation}
Собственная частота:
	\begin{equation}
	\tilde{\omega}_0 \approx \frac{1}{\sqrt{\bar{\epsilon}}+\sqrt{\bar{\epsilon}-1} \cos 2\bar{\chi}}. 
	\label{eq:angle_dispersion_mu1_omega}
	\end{equation}
Это выражение точнее ранее полученного приближения (уравнение (10) в \cite{Timofeev2016t}).
Тем не менее, приближение остается в использовании единого угла $\bar{\chi}$ между оптическими осями холестерика и СПАЗ,
поскольку направления электрической и магнитной поляризаций в общем случае различны.
Это не соответствует условию на поверхности СПАЗ.
Ненулевой угол отличия (согласования) поляризаций $\psi$ приводит к появлению нелокальной компоненты ХОТС \cite{Becchi_Oldano2004}.
В результате состояние 
становится резонансом \hide{leaky mode, or just resonance}
и приобретает конечное время релаксации $\tau_\psi$.

\section*{Спектральное проявление и время релаксации ХОТС}

По определению, время релаксации колебательного состояния есть отношение запасенной энергии $\mathcal{E}$ к мощности потерь $P$, взятой с положительным знаком:
	\begin{equation}
	\frac{\tau}{2} 
	= \frac{\mathcal{E}}{P}
	= \frac{\mbox{Energy stored}}{\mbox{Power of leakage}}.
	\label{eq:tau_def}
	\end{equation}
Здесь $\tau$ -- время релаксации амплитуды, которое вдвое больше энергетического времени релаксации.
Поскольку вглубь идеального СПАЗ поле не проникает, энергия ХОТС есть энергия дифрагирующей в холестерике волны.
Найдем эту энергию, проинтегрировав по пространству ее плотность $W = [\vec{E}\vec{D}+\vec{H}\vec{B}]/8\pi$ в Гауссовой системе единиц.
Для бегущей волны средние за период плотности
электрической и магнитной составляющих энергии равны, поскольку
$ \overline{\vec{E}\vec{D}} = \overline{\vec{H}\vec{B}}$.
При этом $\overline{|E^2|} = |E_0^2|/2$, где $E_{0}$ -- амплитуда волны.
Поэтому 
$\bar{W} = \bar{\epsilon} |E_{0}^2|/8 \pi$.
Стоячую волну в слое образуют две волны одинаковой амплитуды, бегущие в противоположных направлениях.
Их конструктивная интерференция компенсируется деструктивной и плотности энергии складываются
$\bar{W}_0 = \epsilon_h |E_0^2|/4 \pi$.
Интегрирование дает:
	\begin{equation}
	\mathcal{E} = \int_{0}^{\infty} \bar{W}_0 \exp(2 i n_f \cdot 2\pi z/\lambda_0) dz = \frac{\bar{W}_0 \lambda_0}{4\pi\delta \sin 2\bar{\chi}}.
	\end{equation}
Здесь $\lambda_0 = \tilde{\lambda}_0 p$ определено формулой (\ref{eq:angle_dispersion}).
Мощность потерь пропорциональна скорости потока $c / \sqrt{\bar{\epsilon}}$ и плотности энергии
$\bar{W}_\psi = \bar{\epsilon} |E_\psi^2|/8 \pi$, уносимой бегущей волной.
Напряженность определяется граничными условиями:
% \hide{круговая вдвое больше линейной?}
	\begin{equation}
	E_\psi = 2 E_0 \sin (\psi/2). 
	\label{eq:J_psi}
	\end{equation}
В итоге \cite{Becchi_Oldano2004}
	\begin{equation}
	\tau_\psi = 
	\frac{\lambda_f}{4\pi c}
	\frac{1}{\sin^2 (\psi/2)}. 
	\label{eq:t_psi}
	\end{equation}

Рассмотрим слой холестерика конечной толщины $L$,
ограниченный средой с проницаемостью $\bar{\epsilon}_g = \sqrt{\epsilon_o \epsilon_e}$.
Вместо (\ref{eq:J_psi}) напряженность потерь на краю холестерика будет, согласно (\ref{eq:J_eigen}),
	\begin{equation}
	E_L = 2 E_0 \exp \left( - 
	\frac{2\pi L}{\lambda_f}
	% \frac{\pi L \delta_\epsilon \sin 2\bar{\chi}}{\lambda_0}
	 \right). 
	\end{equation}
И соответствующее время релаксации
	\begin{equation}
	\tau_L = 
	\frac{\lambda_f}{4\pi c}
	% \frac{\lambda_0}{2\pi c \delta_\epsilon \sin 2\bar{\chi}}
	\exp \left(
	\frac{4\pi L}{\lambda_f}
	% \frac{2\pi L \delta_\epsilon \sin 2\bar{\chi}}{\lambda_0}
	\right),
	\label{eq:t_L}
	\end{equation}
что согласуется с выражениями, полученными в \cite{Becchi_Oldano2004,Belyakov2011}.

\subsection*{Теория связанных мод во временной области}

Пусть неидеальное локализованное состояние (резонанс)
определяется при помощи собственной частоты $\omega_0$
и $N$ независимых факторов утраты энергии 
с временами релаксации $\tau_\ell, \; \ell=1...N$.
Эти факторы принято называть портами (каналами, континуумами, волноводами).
Таким образом, через $N+1$ действительных величин
неявно учитываются все геометрические и материальные детали структуры.
Согласно ВТСМ
\cite{Joannopoulos2008bk,Manolatou1999}
резонанс описывается комплексной амплитудой $A$.
Эту теорию не следует путать с пространственной теорией связанных мод, рассматривающей амплитуды распространяющихся волн в зависимости от пространственной координаты \cite{Kogelnik1969,Pierce1954}.
Резонанс проявляется через амплитуды 
$s_{\ell\pm}$
потоков входящей и выходящей энергии:
	\begin{eqnarray}
	\frac{dA}{dt}=
	-i\omega_0 A
	-\sum_{\ell=1}^N \frac{A}{\tau_\ell}
	+\sum_{\ell=1}^N \sqrt{\frac{2}{\tau_\ell}} s_{\ell +}, \nonumber \\
	s_{\ell -} = - s_{\ell +} + \sqrt{\frac{2}{\tau_\ell}} A.
	\label{eq:CMT}
	\end{eqnarray}
Возбуждение через один из портов $s_{\ell+}=s_0 \exp{(-i\omega t)}$ дает амплитуду
	\begin{equation}
	A_\ell(\omega) = \frac{\sqrt{\frac{2}{\tau_\ell}} }
	{i(\omega-\omega_0) +
	\sum_{\ell=1}^N {\frac{1}{\tau_\ell}}} s_{\ell+}.
	\label{eq:Amplitude}
	\end{equation}
Амплитуды отражения из порта $\ell$ в порт $\ell'$
образуют матрицу рассеяния
	\begin{equation}
	r_{\ell\ell'} = \frac{s_{\ell'-}}{s_{\ell+}} = 
	-\hat{\delta}_{\ell\ell'} + \frac{\sqrt{\frac{2}{\tau_\ell}}\sqrt{\frac{2}{\tau_\ell'}}}
	{i(\omega-\omega_0) +
	\sum_{\ell''=1}^N {\frac{1}{\tau_\ell''}}},
	\label{eq:SM}
	\end{equation}
где $\hat{\delta}_{\ell\ell'}$ -- символ Кронекера.
Им соответствуют спектральные пики в виде Лоренцевых контуров
с шириной на полувысоте \hide{FWHM}
	\begin{equation}
	2\gamma = 2\sum_{\ell=1}^N {\frac{1}{\tau_\ell}}.
	\label{eq:FWHM_omega}
	\end{equation}
При переходе от циклической частоты $\omega = 2\pi\nu$ к частоте $\nu$, 
ширина пика (\ref{eq:FWHM_omega}) сужается в $2\pi$ раз:
	\begin{equation}
	\Delta\nu =  \frac{\gamma}{\pi} = \frac{1}{\pi}\sum_{\ell=1}^N {\frac{1}{\tau_\ell}}.
	\label{eq:FWHM}
	\end{equation}
Этого достаточно, чтобы описать спектральное проявление состояния при помощи ВТСМ.
\begin{figure*}[htbp]
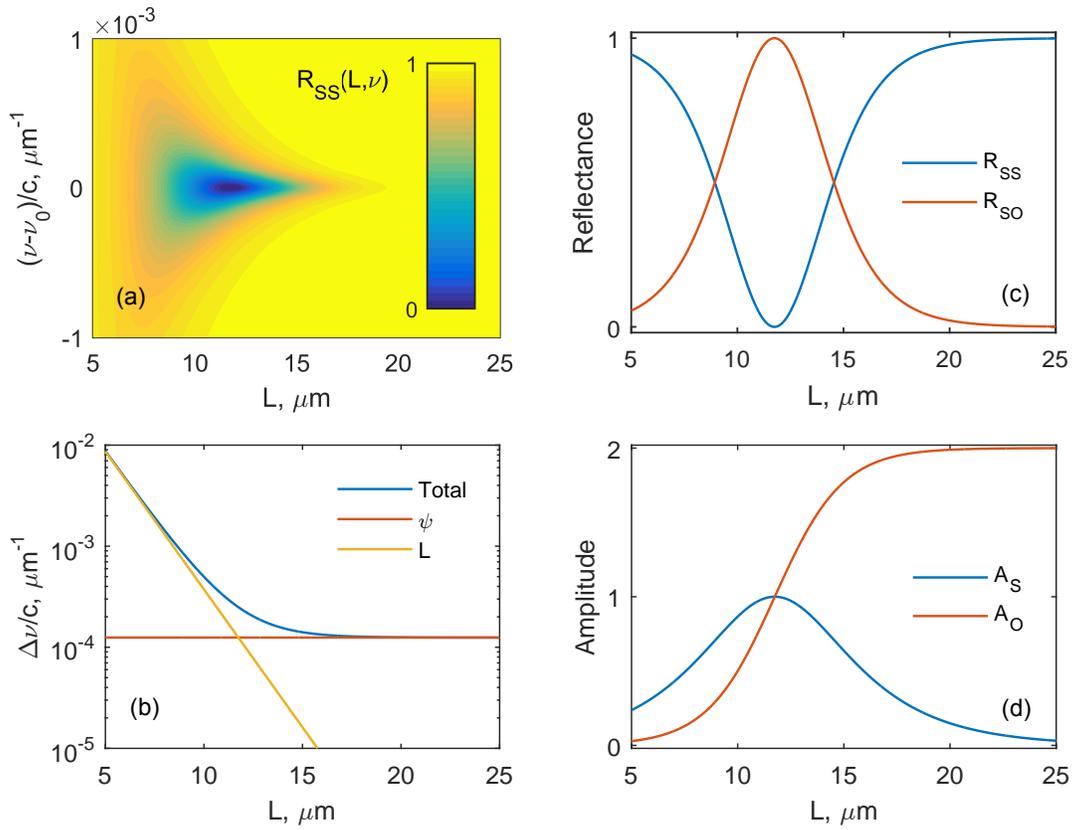
%[htbp]
\includegraphics[scale=0.91]{F4a.pdf}
\includegraphics[scale=0.91]{F4c.pdf}
\includegraphics[scale=0.91]{F4b.pdf}
\includegraphics[scale=0.91]{F4d.pdf}
\caption{%
Эффект Коппа-Генака.
Насыщение ширины линии
при увеличении толщины холестерического слоя.
Смена поляризации оптимально возбуждающего излучения.
(a) Спектр отражения, уравнение (\ref{eq:SM}).
(b) Ширина спектрального провала, уравнение (\ref{eq:FWHM}).
Шаг кручения холестерика $p = 1\mu m$,
электрическая анизотропия $\delta_\epsilon = 0.1$,
магнитная анизотропия отсутствует $\delta_\mu = 0$.
(с) Отражение на резонансной частоте $\omega = \omega_0$, уравнение (\ref{eq:SM_max}).
(в) Амплитуда состояния на резонансной частоте $\omega = \omega_0$, уравнение (\ref{eq:Amplitude}).
% Отличия при расчетах методами ВТСМ и Берремана в масштабе графика не видны.
}
\label{Fig3}
\end{figure*}

В предыдущем разделе было детально описано
идеальное ХОТС, собственная частота которого
дается уравнением (\ref{eq:angle_dispersion}).
Отклонения от идеального состояния,
обусловленные углом $\psi$ и толщиной слоя $L$, 
были выражены
при помощи времен релаксации (\ref{eq:t_psi}) и (\ref{eq:t_L}),
соответственно.
Этого достаточно, чтобы полностью описать спектральное проявление ХОТС при помощи ВТСМ.

% Далее без потери общности условимся рассматривать вращение директора по правому винту.
Порт связи через угол $\psi$ имеет круговую поляризацию,
противоположную ($O$ -- opposite) кручению холестерика,
а через конечную толщину холестерика $L$ -- совпадающую круговую поляризацию ($S$ -- same).
На резонансной частоте $\omega = \omega_0$ уравнение (\ref{eq:SM}) дает матрицу отражений:
	\begin{equation}
	\hat{R} = 
	\left[\begin{array} {cc}
	r_{SS}^2 		& r_{SO}^2 \\ 
	r_{OS}^2 		& r_{OO}^2 \\ 
	\end{array} \right] =
	\left[\begin{array} {cc}
	\cos^2 2\theta		& \sin^2 2\theta \\ 
	\sin^2 2\theta		& \cos^2 2\theta \\ 
	\end{array} \right],
	\label{eq:SM_max}
	\end{equation}
где $\tan \theta = \tau_\psi / \tau_L$.
Эта матрица удовлетворяет закону сохранения энергии.
Максимум недиагональных отражений соответствует кроссоверу.
Он наступает при выравнивании времен релаксации $\theta=\pi/4$.
Приравнивая выражения для  времен релаксации (\ref{eq:t_psi}) и (\ref{eq:t_L}),
получаем аналитическое выражение для длины кроссовера $L_c$:
	\begin{equation}
	\exp(
	\frac{4\pi L_c}{\lambda_f}
	) =
	\frac{1}{\sin^2 (\psi/2)},
	\end{equation}
	\begin{equation}
	L_c = 
	\frac{\lambda_f}{4\pi}
	% \frac{\lambda_0}{2\pi \delta_\epsilon \sin 2\bar{\chi}}
	|\log{\sin^2 (\psi/2)}|. 
	\label{eq:L_crossover}
	\end{equation}
В \cite{Kopp2003_comment} справедливо отмечается, что эту длину трудно измерить в дефекте кручения холестерика
из-за большого $\tau_\psi$ и высокой точности, предъявляемой к эксперименту.
В случае ХОТС величина $\tau_\psi$ становится существенно больше
за счет несовершенства СПАЗ и длина кросовера сокращается,
что должно упростить ее экспериментальное измерение.

\subsection*{Границы применимости ВТСМ}

Применение ВТСМ к решению задачи о связи локализованного состояния с волноводами,
накладывает известные ограничения:
линейность, инвариантность во времени параметров структуры,
сохранение энергии, обратимость потока энергии во времени и слабая связь \cite{Joannopoulos2008bk}.
% weak coupling, linearity, time-invariance (i.e., the materials/geometry don’t change over time), conservation of energy, and time-reversal invariance
Выбранная модель заведомо удовлетворяет первым четырем условиям.
Пятое условие содержит в себе два требования.
Во-первых, решение, получаемое из уравнения (\ref{eq:CMT}),
не учитывает поправки во втором порядке по малому параметру,
равному отношению периода колебаний к времени релаксации
\cite{Haus1984bk}.
Во-вторых, дисперсия волноводов должна быть незначительной на частотном интервале лоренцева контура. 
При достаточно большом времени релаксации ширина лоренцева контура стремится к нулю
и второе требование выполняется автоматически \cite{Joannopoulos2008bk}.

В данном случае холестерик служит волноводом для волн с противоположным знаком круговой поляризации. В середине стоп-зоны дисперсия умеренна. 
Будем полагать, что система линейных дифференциальных уравнений (\ref{eq:CMT}) удовлетворительно описывает модель при $L\ge 5p$ и $\delta_\epsilon=0.1$.

\subsection*{Численные результаты}

Рис.~\ref{Fig3} показывает, что для ХОТС имеет место эффект Коппа-Генака.
Для определенности мы положили шаг геликоида $p=1$ мкм.
При указанных на рисунке параметрах
собственная частота $\nu_0/c=1\mu\mbox{m}^{-1}$ достигается при угле
$\phi = \bar{\chi} = 43^\circ$,
угол несогласования $\psi = 4^\circ$.
Длина кроссовера $L_c=11.75 \mu\mbox{m}$ согласуется с уравнением (\ref{eq:L_crossover}).
При существенно б\'{о}льших толщинах слоя холестерика $L\gg L_c$
ширина линии $\Delta\nu_\infty/c \approx 1.25\cdot 10^{-4} \mu\mbox{m}^{-1}$,
добротность выходит на насыщение $Q_\infty=\nu_0 /\Delta\nu_\infty \approx 8000$,
а при кроссовере $L = L_c$ добротность вдвое меньше.

Рис.~\ref{Fig3},а показывает, что при кроссовере $L = L_c$ 
структура отражает свет как обыкновенное зеркало, 
меняя знак круговой поляризации на противоположный.
С ростом толщины слоя холестерика $L\gg L_c$
его поляризационные свойства возвращаются к свойствам СПАЗ при $L=0$:
знак круговой поляризации при отражении перестает меняться.
Рис.~\ref{Fig3},б показывает, что при кроссовере $L = L_c$ 
правая и левая круговые поляризации одинаково эффективно возбуждают ХОТС.
Следовательно, наиболее эффективное возбуждение наступает при
суперпозиции этих поляризаций с равными амплитудами и определенной разностью фаз.
Эта суперпозиция есть линейно поляризованное излучение.
Ортогонально поляризованное излучение не возбуждает ХОТС.
С ростом толщины слоя холестерика $L\gg L_c$
возбуждение возможно только при помощи света круговой поляризации противоположного знака,
поскольку свет совпадающей круговой поляризации полностью отражается холестериком.

\section*{Заключение}

Хиральное оптическое таммовское состояние 
может считаться строго локализованным 
только когда анизотропии электрической и магнитной проницаемостей
равны по величине,
а оси соответствующих тензоров совпадают.
В противном случае ХОТС проявляется
в виде резонансов поляризованного отражения
с двумя константами релаксации,
определяемыми различием проницаемостей
и толщиной слоя холестерика.
% Для излучения с круговой поляризацией того же знака, что и геликоид холестерика,
% связь экспоненциально ослабляется с толщиной холестерического слоя,
% поскольку частота ХОТС находится в стоп-зоне.
% Излучение с круговой поляризацией противоположного знака
% безпрепятственно проходит сквозь холестерический слой,
% и может возбудить ХОТС при отражении от сохраняющего хиральность зеркала.
ХОТС тесно связано с дефектом кручения холестерика,
и проявляет аналог кроссовер-эффекта Коппа-Генака.
В рамках временной теории связанных мод найдена матрица рассеяния,
описывающая спектральные пики лоренцевыми контурами.
Впервые найдена формула для кроссовер-толщины холестерика,
справедливая как для ХОТС, так и для дефекта кручения.
Аналитическое описание согласуется с прямым численным рассчетом.

% \subsection*{Термины, сокращения}

% \begin{itemize}
% \item ХОТС = хиральное ОТС
% \item состояние (чаще поверхностное) = оптическая мода (чаще дефектная)
% \item локализованное = ограниченное = bound
% \item время жизни состояния = время релаксации
% \item согласование фазы = условие положительной обратной связи
% \item поляризационная устойчивость = класс состояний в континууме, связанных за счет обращения в ноль коэффициента связи
% \item холестерик = ХЖК
% \item геликоид = винт = пространственная спираль
% \item дефект кручения = твист-дефект
% \item круговая дифракция Брэгга = селективное отражение = circular Bragg phenomena
% \item СПАЗ = handedness preserving mirror = handed mirror = hirror = СХЗ = СПАЗ
% \end{itemize}

\bibliography{library}